\documentclass[12pt]{spieman}  
\usepackage{amsmath,amsfonts,amssymb}
\usepackage{graphicx}
\usepackage{setspace}
\usepackage{tocloft}
\usepackage{longtable}

\title{Characterizing Earth analogs may require a moderate or high-resolution spectrograph}

\author[a,*]{Jean-Baptiste Ruffio}

\author[b]{Sarah Steiger}
\author[c]{Corey Spohn}
\author[d,e]{Bruce Macintosh}
\author[f,g]{Dimitri Mawet}
\author[b]{Laurent Pueyo}
\author[g]{Bertrand Mennesson}
\author[a]{Beck	Dacus}
\author[d]{Nicole Wolff}
\author[h]{Tyler D. Robinson}
\author[g]{Renyu Hu}
\author[b]{Kielan Hoch}
\author[a]{Quinn M. Konopacky}
\author[b]{Marshall D. Perrin}
\author[i]{Dmitry Savransky}
\author[c]{Michael W. McElwain}
\author[a]{Shelley A. Wright}
\author[j]{Ji Wang} 
\author[g]{Pin Chen}

\affil[a]{Department of Astronomy \& Astrophysics, University of California San Diego, La Jolla, CA, USA}
\affil[b]{Space Telescope Science Institute, 3700 San Martin Drive, Baltimore, MD 21218, USA}
\affil[c]{NASA Goddard Space Flight Center, Greenbelt, MD 20771, USA}
\affil[d]{Department of Astronomy \& Astrophysics, University of California, Santa Cruz, CA95064, USA}
\affil[e]{University of California Observatories, 1156 High Street, Santa Cruz, CA 95064, USA}
\affil[f]{Department of Astronomy, California Institute of Technology, Pasadena, CA 91125, USA}
\affil[g]{Jet Propulsion Laboratory, California Institute of Technology, 4800 Oak Grove Drive, Pasadena, CA 91109, USA}
\affil[h]{Lunar \& Planetary Laboratory, University of Arizona, Tucson, AZ, USA}
\affil[i]{Sibley School of Mechanical and Aerospace Engineering, Cornell University, Ithaca, NY, 14853}
\affil[j]{Department of Astronomy, The Ohio State University, 4055 McPherson Laboratory, 140 West 18th Avenue, Columbus, OH 43210, USA}

\cftpagenumbersoff{figure}
\cftpagenumbersoff{table} 
\begin{document} 
\maketitle

\begin{abstract}
A primary goal of the Habitable Worlds Observatory (HWO) is to detect and measure the abundance of biosignature molecules, such as water (H$_2$O) and oxygen (O$_2$), in the atmosphere of Earth analogs. This is expected to require deep spectroscopic observations lasting hundreds of hours per planet.
In this context, it is essential to optimize the spectral resolution of the spectrograph to both maximize the number of planets that can be studied over the lifetime of the mission, and also to reduce the risks of false detections.
The purpose of this work is to provide a framework to explore the spectral resolution design trade-space for HWO. This framework must be valid and comparable across all spectral resolutions from low ($R<100$) to high resolutions ($R>10,000$), and account for the spectral correlation of the residual starlight (i.e., speckle noise chromaticity).
Leveraging the concept of ``template matching'', we develop a simulation toolkit based on the Python package \texttt{EXOSIMS} to compute the detection significance of planets and molecules. 
We then simulate observations of Earth analogs around 164 stars using representative mission parameters to explore the effects of the detector noise and the correlated speckle noise floor.
Our findings suggest that a moderate or high resolution spectrograph ($R>1,000$) will provide higher sensitivity to critical molecules compared to a low resolution spectroscopy mode (e.g., $R\sim140$). The correlated speckle noise may also entirely suppress our ability to detect bio-signatures at low spectral resolutions.
We conclude that a more comprehensive study combined with detailed models of its stability, and other sources of correlated noise, is necessary to fully explore the trade space of spectral resolution and detectability of key species.
\end{abstract}


{\noindent \footnotesize\textbf{*}Jean-Baptiste Ruffio,  \linkable{jruffio@ucsd.edu} }

\begin{spacing}{2}

\section{Introduction}
\label{sect:intro}  

The Habitable Worlds Observatory (HWO) is NASA Astrophysics' next planned flagship mission\cite{Feinberg2024SPIE13092E..1NF}. Several studies in the literature have contributed to the exploration of the spectral resolution trade-space for HWO, some at low-spectral resolution ($R<300$), and some at moderate to high-spectral resolution spectroscopy ($R>1,000$; MHRS).

\subsection{Low spectral resolution}
Design of precursor concepts \cite{Gaudi2020arXiv200106683G, LUVOIR2019arXiv191206219T} and exploratory analytic cases (EAC)\cite{Feinberg2026arXiv260111803F} for HWO have typically assumed a low spectral resolution spectrograph ($R\sim140$) for the detection of molecular signatures.
This choice has been primarily motivated by the low photon count rate of Earth analogs in reflected light ($\sim10^{-10}$ planet-to-star flux ratios) and the impact of detector noise when increasing the spectral resolution due to light being spread over more pixels. In this work, we will revisit these points and discuss the extent to which planet flux and detector noise might impact planet and molecule detection.

At low spectral resolution, mock atmospheric inferences\cite{Feng2018AJ....155..200F,Brandt2014PNAS..11113278B,Damiano2022AJ....163..299D,Damiano2023AJ....166..157D,Zhang2024PASP..136e4401Z} have suggested that O$_2$ and H$_2$O could be meaningfully detected assuming a signal-to-noise ratio (S/N) per spectral bin of 10 to 20 at $R=140$. 
However, a significant caveat is that the presence of correlated noise is usually neglected. If not accurately modeled, correlated residual speckles will bias atmospheric inference, especially at low spectral resolution \cite{Greco2016ApJ...833..134G}. 
Indeed, spectral features cannot be effectively disentangled from the speckle noise if they have similar amplitude and wavelength scales.
Important sources of correlated noise include: the magnification of the stellar point spread function (PSF) with wavelength\cite{Brandt2014PNAS..11113278B}, wavefront variability\cite{McElwain2023PASP..135e8001M}, any source of fringing\cite{Horstman2024SPIE13096E..2EH,Gasman2023A&A...673A.102G}, or spectral extraction systematics\cite{Law2023AJ....166...45L,Ruffio2024AJ....168...73R}. Even systematics in atmosphere models can effectively create correlated noise residuals\cite{Czekala2015}.
Each source of correlated noise can usually be characterized by a typical correlation length scale in the spectral direction, an amplitude, and a timescale of its variability. It is essential to characterize these sources of noise as they will impact the ability to study exoplanet atmospheres.

\subsection{Moderate to high-spectral resolution}
MHRS can help mitigate the impact of correlated noise. 
At high enough spectral resolution, it is for example possible to high-pass filter the spectra to remove the correlated noise while retaining the distinct spectral features of molecules if they are sufficiently resolved. Molecules may then be detected using template matching techniques like cross correlation function (CCF)\cite{Konopacky2013Sci...339.1398K, Hoeijmakers2018A&A...617A.144H,Ruffio2019AJ....158..200R}. 

MHRS has proven to be a powerful technique for atmospheric characterization of exoplanets with ground-based telescopes thanks to its relative insensitivity to PSF variability\cite{Snellen2025arXiv250508926S}. 
For example, it has enabled population-level atmospheric characterization of C/O and metallicity\cite{Hoch2023AJ....166...85H,Xuan2024ApJ...970...71X,Wang2025ApJ...981..138W}.
The success of MHRS techniques have also been demonstrated in space with JWST/NIRSpec, a $R\sim2,700$ integral field spectrograph not designed for high-contrast observing, showing improved high-contrast sensitivity at small projected separation ($<1^{\prime\prime}$) compared to imaging \cite{Ruffio2024AJ....168...73R}.

A number of studies have explored the potential of MHRS for exoplanet characterization with future facilities\cite{Landman2023A&A...675A.157L,Bidot2024A&A...682A..10B,Hardegree2025AJ....169..171H, Wang2017AJ....153..183W} including in the context of HWO precursor concepts\cite{Wang2017SPIE10400E..0ZW,Wang2018JATIS...4c5001W}. These studies typically rely on high-pass filtering and CCF. High-pass filtering is often used because it is the most direct way to mitigate the correlated noise by simply removing it, but this means paying the price of the loss of the planet continuum. Although high-pass filtering spectra does not generally prevent the characterization of exoplanet atmospheres, it does remove the continuum of the planet spectrum.
Fortunately, high-pass filtering would not be required with a very stable PSF. Classical PSF subtraction techniques do work with MHRS when the PSF is sufficiently stable; for example with JWST\cite{Ruffio2024AJ....168...73R}.
It is therefore not necessary to assume high-pass filtered spectra for MHRS with an ultra-stable telescope like HWO. This means that the potential of MHRS should be explored while retaining the planet continuum, which is important for planet detection and characterization. 

A significant concern when increasing the spectral resolution for HWO is the increased impact of detector noise due to the spreading of the light over a larger number of pixels. However, increasing the spectral resolution also increases the information content in the spectrum and could provide more robust biosignature detections. Additionally, MHRS is likely to be more robust to astrophysical false positives such as non-uniformities in exozodiacal dust distribution or background sources thanks to template matching. Critically, it could also provide valuable additional information for the study of giant planets including radial velocities, spins, and other trace molecules that are only accessible at higher spectral resolution.

\subsection{Outline}
There is a clear need for more comprehensive studies of spectral resolution for HWO. Such studies need to not only enable direct comparisons of performance between any spectral resolution ($R\sim140$ to $R>10,000$), but also include a framework to account for the correlated noise.

In Section~\ref{sect:tmsnr}, we first define the statistical framework to compute planet and molecular S/Ns based on template matching. 
Then, we describe the simulation framework based on the Python package \texttt{EXOSIMS} in Section~\ref{sect:simus}. The results of some initial simulations are presented in Section~\ref{sect:results}. Finally, we discuss our findings in Section~\ref{sect:discussion} and conclude in Section~\ref{sect:conclusion}.

\section{Template matching S/N}
\label{sect:tmsnr}

Template matching, or matched filtering, provides a convenient framework to define the signal-to-noise ratio (S/N) of a planet detection. This maximum-likelihood based framework satisfies the needs highlighted previously: it is valid at any spectral resolution and includes a model of the noise covariance assuming Gaussian noise. It can also be used to evaluate the planet detection S/N as well as the significance of the detection of any individual molecule in the atmosphere.

\subsection{Planet detection S/N}
\label{sect:pldetec}

We first define the statistical framework used to fit a planet model to noisy data.
An observed spectrum  ($\mathbf{d}=[d_1,d_2,\dots]^\top$; the data) can be decomposed into noise and model components such that: 
\begin{equation}
    \mathbf{d} = a\mathbf{m} + \mathbf{n},
\end{equation}
where $\mathbf{m}=[m_1,m_2,\dots]^\top$ is the planet model spectrum scaled by the planet flux $a$. $\mathbf{n} \sim \mathcal{N}(\boldsymbol{0}, \boldsymbol{\Sigma})$ is random vector following a zero-mean multivariate normal distribution with covariance matrix $\boldsymbol{\Sigma}$. $\mathbf{d}$ is the extracted spectrum of the planet, with the elements $d_i$ being the photon counts in each spectral bin.

The best-fit planet flux $\tilde{a}$ and its associated standard deviation $\tilde{a}_{\mathrm{err}}$ can be derived via maximum likelihood estimation (e.g., [\citenum{Ruffio2017ApJ...842...14R}]):

\begin{align}
    \tilde{a} &= \frac{\mathbf{d}^\top \boldsymbol{\Sigma}^{-1} \mathbf{m}}{\mathbf{m}^\top \boldsymbol{\Sigma}^{-1} \mathbf{m}} \\
    \tilde{a}_{\mathrm{err}} &= \frac{1}{\sqrt{\mathbf{m}^\top \boldsymbol{\Sigma}^{-1} \mathbf{m}}}
\end{align}

Consequently, the signal-to-noise ratio is given by:

\begin{equation}
    \mathrm{S/N} = \frac{\tilde{a}}{\tilde{a}_{\mathrm{err}}} = \frac{\mathbf{d}^\top \boldsymbol{\Sigma}^{-1} \mathbf{m}}{\sqrt{\mathbf{m}^\top \boldsymbol{\Sigma}^{-1} \mathbf{m}}}
\end{equation}

In the context of S/N simulations, meaning in the absence of real observations $\mathbf{d}$, the S/N can be further simplified by calculating the average expected S/N (over many hypothetical observations) as a function of the noise and model only. Following the prescription in [\citenum{Landman2023A&A...675A.157L}], but retaining the covariance in the S/N definition, we get:
\begin{equation}
    \mathrm{S/N} = \sqrt{\mathbf{m}^\top \boldsymbol{\Sigma}^{-1} \mathbf{m}}
    \label{eq:tmsnr}
\end{equation}
where we have assumed that the model ($\mathbf{m}$) is scaled such that the planet flux ``$a=1$''. This means that $m_i$ is already calculated to be the expected photon count of the planet in each spectral bin.

This idealized model represents the fundamental limit for the detection significance assuming perfect data reduction other than what is modeled by the covariance matrix. In other words, it achieves the Cramer-Rao bound when the models of the planet and the noise are accurate. We note that real data will always feature specific challenges not modeled in Equation~\ref{eq:tmsnr} that might require complex data analysis to address. These data challenges might include spectral fringing from transmissive optics, detector persistence, curvature of the spectral trace on the detector, etc. However, when these issues are adequately calibrated, they would not necessarily limit the performance of the system and can therefore be neglected in simplified simulations.

\subsection{Molecule detection S/N}
\label{sect:moldetec}

\begin{figure}
\begin{center}
\begin{tabular}{c}
    \includegraphics[width=\linewidth]{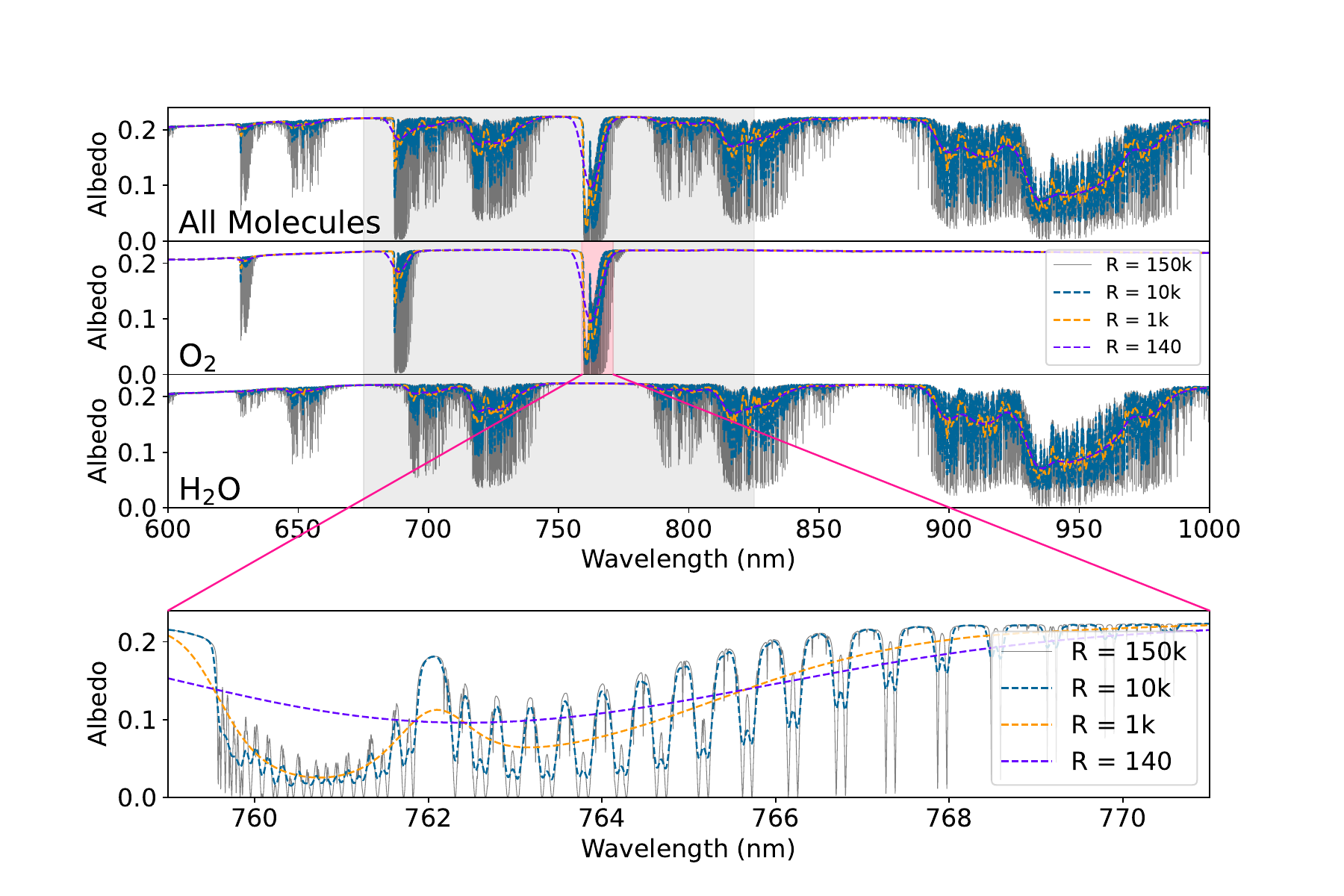}
\end{tabular}
\end{center}
\caption 
{ \label{fig:spectra}
Spectral dependence of the albedo of an Earth analog\cite{Meadows1996JGR...101.4595M,Robinson2011AsBio..11..393R} as a function of spectral resolution ($R$). (Top) The top panel shows the combined spectrum, while the middle and bottom panels highlight the contributions of individual molecules, namely O$_2$ and H$_2$O. The light grey area corresponds to a 20\% bandpass centered at 750~nm. (Bottom) A zoomed-in region (pink highlight) on top of the oxygen spectral feature at 762~nm.} 
\end{figure}

Regardless of spectral resolution, the detection of a molecule can be seen as a measure of how much a spectrum deviates from a ``flat line''. Fig.~\ref{fig:spectra} shows an example of exo-Earth molecular templates for O$_2$ and H$_2$O.
The high-resolution spectral models were computed using the Spectral Mapping Atmospheric Radiative Transfer (SMART) model (developed by D.~Crisp)\cite{Meadows1996JGR...101.4595M}. Clear-sky, low-cloud, and high-cloud scenarios were developed according to earlier, validated models \cite{Robinson2011AsBio..11..393R}. These scenarios were linearly combined to model a planet with a composite surface that is 50\% cloud-free, 25\% covered in low-altitude clouds, and 25\% in high-altitude clouds following the prescription in Ref.\cite{Robinson2011AsBio..11..393R}. The overall albedo is also normalized to yield a 0.2 average albedo in the $675$--$825$\,nm range.
The models include gaseous absorption due to water vapor, carbon dioxide, ozone, methane, carbon monoxide, nitrous oxide, and molecular oxygen. The resolution of these simulations varied with wavelength so as to resolve (with at least eight spectral points) every line for these gases in the HITRAN 2020 database \cite{Gordon2022JQSRT.27707949G}.

For a spectrum with sparse absorption features, we define the envelope of the spectrum as the smooth curve that traces the upper bound of the spectrum while ignoring localized absorption features.
After subtracting the envelope of the molecular template ($\mathbf{m}_{\mathrm{mol}}=\mathbf{m}-$``envelope''), we can measure the molecular detection S/N using Eq.~\ref{eq:tmsnr}, such that: 

\begin{equation}
    \mathrm{S/N}_{\mathrm{mol}} = \sqrt{\mathbf{m}_{\mathrm{mol}}^\top \boldsymbol{\Sigma}^{-1} \mathbf{m}_{\mathrm{mol}}}
    \label{eq:molsnr}
\end{equation}
This S/N measures how much the spectrum deviates from the smooth continuum following the molecular template.

\subsection{Modeling the covariance}
\label{sect:cov}

It is essential to model the spectral covariance for any S/N calculation or atmospheric inference\cite{Greco2016ApJ...833..134G}. Neglecting the covariance can lead to significant biases in the estimated parameters and S/N values. This is expected to be a significant source of uncertainties in existing studies for exoEarth and biosignature detection. While some studies have included correlated noise like Ref.\cite{Lupu2016AJ....152..217L}, which runs simulated retrievals of reflected-light gas giant atmospheric retrievals, correlated noise has often been neglected.
In the following, we solely consider the residual correlated noise after speckle subtraction, which can be different from the typical correlations of the raw speckles.
A covariance is generally defined with a uncorrelated noise component and correlated component, such as:

\begin{equation}
    {\Sigma} = {\Sigma}_{\mathrm{uncorr}}+{\Sigma}_{\mathrm{corr}}
\end{equation}

In the following, ${\Sigma}_{\mathrm{uncorr}}$ is a diagonal matrix with diagonal elements $\sigma_i^2$ representing the uncorrelated noise in each spectral bin including the combined total photon noise and detector noise. ${\Sigma}_{\mathrm{corr}}$ represents the correlated speckle noise residuals in the extracted planet spectrum after PSF subtraction. 
In the following, we consider a simple model for ${\Sigma}_{\mathrm{corr}}$ with a Gaussian kernel profile \cite{Greco2016ApJ...833..134G}. In the absence of a more detailed model of the correlated noise based on detailed models of the instrument and observatory behavior, a Gaussian profile is sufficient to evaluate the first order impact of the correlated noise. We then have
\begin{equation}
\Sigma_{\mathrm{corr},i,j} = \sigma_{i,\mathrm{corr}} \sigma_{j,\mathrm{corr}}  \exp\left( -\frac{1}{2} \frac{(\lambda_j - \lambda_i)^2}{\Lambda_{\mathrm{corr}}^2} \right)
\end{equation}
where $\sigma_{i,\mathrm{corr}}$ is the standard deviation of the correlated noise. 
$\lambda_i$ is the wavelength of the $i$-th spectral bin.
$\Lambda_{\mathrm{corr}}$ is the length scale of the correlated noise defined as the standard deviation of the Gaussian profile.
In the following, we assume $\sigma_{i,\mathrm{corr}}$ to be proportional to the starlight contamination at the position of the planet ($\zeta_{\mathrm{floor}}$). As an example, there could be a systematic difference in the wavefront errors between two ADI roll angles, which would translate into a correlated systematic noise floor if applying pair-wise PSF subtraction. Any such systematics can be modeled to first order by a PSF subtraction gain $g$, which models our ability to subtract the starlight such that $\sigma_{i,\mathrm{corr}}=g\zeta_i$. 

The correlation length will depend on chromatic behavior of residuals speckles. An upper limit on the correlation length is set by the magnification of the PSF with wavelength. In other words, speckle will move across the position of the planet as the wavelength increases. The corresponding correlation scale is set by (Wolff et al.; in prep)
\begin{equation}
\Lambda_{\mathrm{corr}} = \frac{1.22}{2\sqrt{\ln 2}} \cdot \frac{\lambda^2}{D\times \mathrm{WA}},
\label{eq:corrlen}
\end{equation}
With $D$, the diameter of the observatory primary mirror. $\lambda$ is the wavelength, and $\mathrm{WA}$ is the ``working angle'', meaning the angular separation of the planet in radians.
At $\sim3\lambda/D$ and $\lambda\sim750\,$nm, the correlation length scale is $\Lambda_{\mathrm{corr}}\sim200\,$nm, which is larger than the size of a 20\% spectral bandpass. 

More generally, we can assume that speckles will be highly chromatic for HWO in the dark zone after speckle nulling, which will likely result in a shorter correlation length scale\cite{Wang2017AJ....153..183W}.
Determining this effective correlation scale requires detailed models of the coronagraph system and the spectrograph, which are not available at this time. 
Because of the diversity of correlated noise sources, correlated noise can exist at any spectral resolution. For example, fringing or optical micro-roughness could produce correlated systematics at moderate to high spectral resolutions.
However, we can assume that low order and mid-spatial frequency wavefront variability is likely to dominate in a typical optical system, which is why speckle noise is primarily an issue at low spectral resolutions. 

\begin{figure}
\begin{center}
\begin{tabular}{c}
    \includegraphics[width=\linewidth]{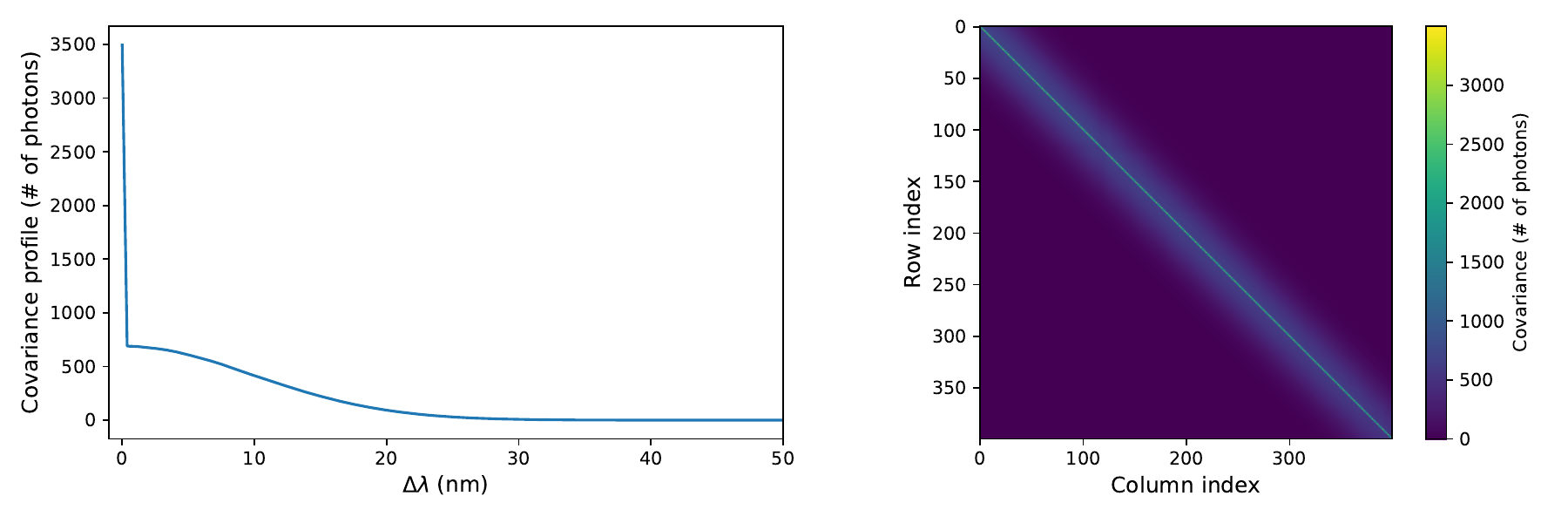}
\end{tabular}
\end{center}
\caption 
{ \label{fig:cov}
Covariance model. (Left) First column of the covariance matrix illustrating the uncorrelated and correlated components of the covariance. (Right) Image of the covariance matrix. A LUVOIR-analog detector noise level with $10^{-10}$ starlight suppression and $10^{-11}$ correlated noise floor were assumed to compute this model. } 
\end{figure}

As a simplified model in this work, we clip the correlation length values given by the PSF magnification (Eq.~\ref{eq:corrlen}) across the bandpass to $\Lambda_{\mathrm{corr,max}}=10\,$nm as a place holder. This is also similar to the typical length scale of the speckle spectra derived from simulations of a notional space telescope with representative wavefront errors\cite{Wang2017AJ....153..183W}.
Future work should revisit this assumption. The covariance model is illustrated in Fig. \ref{fig:cov}. 
In practice, inverting noisy covariances derived empirically from real data can be challenging due their large conditioning numbers. However, the simple noiseless covariance used in this work can be inverted more reliably and the numerical stability is less of a concern.

\subsection{High-pass filtered S/N}
\label{sect:hpfsnr}

At sufficiently high spectral resolution, the correlation length of the residual speckles is typically much larger than the spectral features of the molecules of interest\cite{Landman2023A&A...675A.157L, Ruffio2019AJ....158..200R}. In this case, it is useful to imagine the information content of the planet signal as two components\cite{Landman2023A&A...675A.157L}: 1) the continuum or large-scale features, which corresponds to low spectral frequencies in Fourier space (i.e., low-pass filter), and 2) the small-scale molecular features, which corresponds to high spectral frequencies in Fourier space (i.e., high-pass filter). The role of the covariance is effectively to assign the appropriate level of noise to each spectral resolution, which will be a function of the spectral scale of the correlated noise. 

The S/N of the continuum ($R<\lambda/\Lambda_{\mathrm{corr}}$) is typically limited by the speckle residual systematics. The large-scale correlations mean that oversampling the spectrum does not improve the precision of the estimated continuum, because the spectral bins are no longer independent. When fitting atmospheric models to a noisy spectrum, it is therefore essential to include the covariance in the definition of the $\chi^2$; e.g. $\chi^2=(\mathbf{d}-\mathbf{m})^\top \boldsymbol{\Sigma}^{-1}(\mathbf{d}-\mathbf{m})$. It is not appropriate to only inflate the noise per bin such that ($\sigma_i^2 = \sigma_\mathrm{corr}^2+\sigma_\mathrm{uncorr}^2$) while assuming independent spectral bins with $\chi^2=\sum_i (d_i-m_i)^2/\sigma_i^2$. This approach would overestimate the information content of the data at low spectral resolutions by underestimating the effect of the correlations. However, a common strategy to address this issue is to down-sample the data, by binning the spectrum, such that neighboring bins are mostly independent.

Conversely, the S/N of the small-scale spectral features ($R>\lambda/\Lambda_{\mathrm{corr}}$) are in general not impacted by the large-scale correlated noise. Indeed, the smooth correlated noise cannot explain the higher spectral resolution features, and will therefore not significantly impact our ability to detect molecules at high enough spectral resolution. This point is one of the main reasons why MHRS has proven powerful for measuring atmospheric abundances of directly imaged planets \cite{Hoch2023AJ....166...85H,Xuan2024ApJ...970...71X}. Naturally, the other motivation for MHRS is the higher information content of high-resolution spectra.

We therefore define an S/N for a high-pass filtered spectrum , where the features broader than the correlation length are removed. In this case, the spectrum becomes mostly uncorrelated such that the S/N is given by:

\begin{equation}
    \mathrm{S/N}_{\mathrm{HPF}} = \sqrt{\sum_i m_i^2/\sigma_{i,\mathrm{uncorr}}^2},
    \label{eq:hpfsnr}
\end{equation}
where the residual systematic floor of the speckle subtraction does not impact the S/N. However, this is only true if $\Lambda_{\mathrm{corr}}>>\lambda/R$ so we do not compute the high-pass filtered S/N when the spectral resolution is too low.
For a given correlation length scale $\Lambda_{\mathrm{corr}}$, $\mathrm{S/N}_{\mathrm{HPF}}$ provides an S/N floor that will be independent of $\sigma_{i,\mathrm{corr}}$, and therefore independent of the level of PSF subtraction systematics ($g\zeta_i$). In the following, we use a Gaussian kernel for the high-pass filtering.

\section{Simulation framework}
\label{sect:simus}

 In this section, we define the observatory parameters and Universe assumptions from which $\mathbf{m}$, $\mathbf{m}_{\mathrm{mol}}$, and ${\Sigma}$ are derived. For clarity, the variables and parameters are also summarized in Table~\ref{tab:counts} for the photon count rates, in Table~\ref{tab:para_astro} for the astrophysical parameters, and in Table~\ref{tab:para_obs} for the observatory parameters.

\begin{table}[h!]
\centering
\caption{Total photon count rate (i.e., number of photons per second) in the spectral bandpass (upper case $C$) and count per spectral bin (lower case $c$).}
\begin{tabular}{@{}lp{0.8\linewidth}@{}}
\hline
\textbf{Parameter} & \textbf{Description} \\
\hline
$C_\mathrm{s}$ \& $c_{\mathrm{s},i}$ & Unobstructed starlight without coronagraph.\\
$C_{\mathrm{s},\mathrm{cont}}$ \& $c_{\mathrm{cont},i}$ & Local starlight contamination in the planet spectrum. \\
$C_{\mathrm{s},\mathrm{corr}}$ \& $\sigma_{\mathrm{corr},i}$ & Measure of the level of correlated noise. $\sigma_{\mathrm{corr},i}$ is the standard deviation in phot/s of the correlated noise in spectral bin. \\
$C_\mathrm{z}$ \& $c_\mathrm{z,i}$ & Zodiacal light\\
$C_\mathrm{ez}$ \& $c_\mathrm{ez,i}$ &Exozodiacal light\\
$C_\mathrm{p}$ \& $c_\mathrm{p,i}$ & Planet light\\
$C_\mathrm{dark}$\& $c_\mathrm{dark}$ & Count rate for the dark current\\
$C_\mathrm{RN}$ \& $c_\mathrm{RN}$ & Effective count rate of the read-noise\\
$C_\mathrm{CIC}$ \& $c_\mathrm{CIC}$ & Effective count rate for the clock-induced charges\\
\hline
\end{tabular}
\label{tab:counts}
\end{table}

\begin{table}[h!]
\centering
\caption{Astrophysical parameters. Variables without values are computed within EXOSIMS for each star in the target list.}
\begin{tabular}{@{}llp{0.65\linewidth}@{}}
\hline
\textbf{Parameter} & \textbf{Value/Units} & \textbf{Description} \\
\hline
\multicolumn{3}{@{}l}{\textbf{Stellar}} \\
\# stars & 164 & Number of stars in ExEP HWO Star List 2023\cite{Mamajek2024arXiv240212414M}. \\
$F_s$ & phot/m$^2$/s & Total spectral flux of the star in the observing spectral bandpass. \\
$\mathrm{ZP}_V$ & phot &Vega zero point of the spectrograph bandpass.\\
$L_\mathrm{star}$ & W & Stellar bolometric luminosity.\\
$L_\mathrm{Sun}$ & W & Sun bolometric luminosity.\\

\multicolumn{3}{@{}l}{\textbf{Backgrounds}} \\
$Z$& $23\,$mag/as$^{2}$ & Magnitude of the zodiacal light per angle area. This is assumed to be constant for all stars. \\
$\mathrm{JEZ}_0$& phot/s/m$^2$/as$^2$& Exo-zodiacal light at 1~au based the solar sytem value. \\
$\mathrm{JEZ}$& phot/s/m$^2$/as$^2$& Scaled exo-zodiacal light from Eq.~\ref{eq:JEZ}. \\
 $n_\mathrm{ez}$ & 3&  Scaling factor of the exozodi level.\\
 
\multicolumn{3}{@{}l}{\textbf{Planet}} \\
$R_p$ & 1 $R_\oplus$ & Earth radius \\
$\beta_0$ & $\pi/2$ (half-moon) & Quadrature phase \\
$\phi(\beta_0)$ & $1/\pi$ & Lambertian phase function \\
$d_\mathrm{eed}$ & $(1\,\mathrm{au})\sqrt{\frac{L_\mathrm{star}}{L_\mathrm{Sun}}}$& Earth equivalent distance for a planet with Earth instellation.\\
$p$ & 0.2 & Planet geometric albedo \\

\hline
\end{tabular}
\label{tab:para_astro}
\end{table}

\begin{table}[h!]
\centering
\caption{Observatory parameters.}
\begin{tabular}{@{}lp{0.2\linewidth}p{0.6\linewidth}@{}}
\hline
\textbf{Parameter} & \textbf{Value} & \textbf{Description} \\
\hline
\multicolumn{3}{@{}l}{\textbf{Optics}} \\
$D$ & 7.87 m & Pupil diameter \\
$A$ & $42.8\,\mathrm{m}^2$ & Effective collecting area of the primary mirror.\\
$\tau_\mathrm{opt}$ & 0.532 & Optics transmissions \\

\multicolumn{3}{@{}l}{\textbf{Spectrograph}} \\
$T_{\mathrm{tot}}$ & 400 h/star & Total exposure time per system. \\
$\lambda$ & 750 nm & Central wavelength \\
BW & 20\% & Bandpass \\
$R$ & 20--10,000 & Effective spectral resolution of the spectrograph including the spectral bin broadening.\\
$N_{\mathrm{bin/res}}$& 2 (or 1) & Number of spectral bins per resolution element. \\
$\mathrm{platescale}$& 0.015 as & Spatial pixel dimension on the sky. \\
$N_{\mathrm{pix/bin}}$ & $\sim8$ (or 4) & Number of detector pixels per spectral bins: $N_{\mathrm{bin/res}} \frac{\Omega}{\mathrm{platescale}^2}$ \\
$N_\mathrm{bins}$& $R\times\mathrm{BW}$ & Number of spectral bins in the spectrum. \\
 
\multicolumn{3}{@{}l}{\textbf{Coronagraph}} \\
$\zeta$ & -- & Starlight suppression from USORT OVC coronagraph model.  \\
$\zeta_{\mathrm{floor}}$ &  $10^{-9}$--$10^{-10}$ & Starlight suppression floor; i.e., minimum value of $\zeta$. \\
$g$ & $0.1$--$0.001$ & Post-processing gain. \\
$g\zeta$ & -- & Level of correlated noise with minimum value $g\zeta_{\mathrm{floor}}$. \\
$\Delta\lambda_{\mathrm{corr}}$ & 10 nm & Correlation length scale \\
$\Omega$ & $2.26(\lambda/D)^2$ & Area of the aperture, or core, of the off-axis PSF \\
$\tau_{\mathrm{occ}}$ & e.g. 51\% at 0.1$''$ & Occulter transmission from USORT OVC coronagraph model. Function of projected separation (Fig.~\ref{fig:throughput}). \\
$\tau_{\mathrm{core}}$ & e.g. 25\% at 0.1$''$ & Off-axis core throughput from USORT OVC coronagraph model. Function of projected separation (Fig.~\ref{fig:throughput}). \\

\multicolumn{3}{@{}l}{\textbf{Detector}} \\
$\mathrm{QE}$ & 0.675 & Quantum efficiency \\
$\xi$ & $3 \times 10^{-4}$~phot/s to 0~phot/s & Dark current \\
$\mathrm{RN}$ & $0$ phot/$T_{\mathrm{exp}}$ & Read noise \\
$\mathrm{CIC}$ & $8 \times 10^{-3}$ phot/$T_{\mathrm{exp}}$ to 0 phot/$T_{\mathrm{exp}}$ & Clock-induced charges \\
$T_{\mathrm{exp}}$ & 300~s & Individual exposure time for read-noise and CIC.\\

\hline
\end{tabular}
\label{tab:para_obs}
\end{table}

\subsection{EXOSIMS}
As one of the major mission simulators already used in the context of Roman\cite{Spergel2015arXiv150303757S} and HWO, we leverage the extensive framework provided by \texttt{EXOSIMS} \cite{Savransky2010PASP..122..401S,Savransky2016JATIS...2a1006S, 2016SPIE.9911E..19D, Savransky2017ascl.soft06010S}. \texttt{EXOSIMS} was cross-validated with other exposure time calculators for HWO\cite{Stark2025arXiv250218556S}. We developed a S/N simulation toolkit within the \texttt{EXOSIMS} framework that is valid at any spectral resolution and includes a model of the spectral covariance using the template matching formalism introduced previously. 
A significant advantage of using a framework like \texttt{EXOSIMS} is that it can simulate an observatory as well as a Universe with a star catalog and a population of exoplanets. Our framework is available on Github\footnote{\url{https://github.com/jruffio/EXOSIMS}} as a Git fork from EXOSIMS.

\subsection{Universe assumptions}

We assume a Universe with one Earth analog per star for each of the 164 stars in the ExEP HWO target list that is already part of  \texttt{EXOSIMS}\cite{Mamajek2024arXiv240212414M}. For a set of assumptions, the performance of the observatory will be evaluated from the distribution of S/N values around this mock sample of planets.
We assume all planets have a radius of $R_\mathrm{p}=1\,$Earth radius. We also assume that they  are observed at their maximum elongation in an orbit with similar instellation to Earth (i.e., one earth-equivalent distance; $d_\mathrm{eed}$) and at quadrature phase (i.e., half-disk visible; $\beta_0=\pi/2$). A Lambertian phase function of $\phi(\beta_0=\pi/2)=1/\pi$ is used with a geometric albedo of $p=0.2$ around $750\,$nm. The planet-to-star flux ratio is of the order of $\sim10^{-10}$ but varies with the stellar spectral type; the flux ratio is higher for later type stars. We use the albedo spectrum described in Section~\ref{sect:moldetec} and in Figure \ref{fig:spectra}.
 
In this work, we assume that the stellar spectrum has been subtracted, or modeled, to within the photon noise limit. For integral field spectrographs, this can for example be achieved by deriving an empirical model of the star directly from the speckle field since the planet signal is spatially resolved\cite{Ruffio2024AJ....168...73R}. In practice, the planet signal is detected by looking for deviations from the measured stellar spectrum, but this can be abstracted when the goal is only to estimate the achievable detection S/N of the planet or molecules.

We currently do not include the stellar spectral features in the spectrum of the planet. This is due to the lack of built-in high resolution spectral models for the stars in \texttt{EXOSIMS}. In reflected light, the spectral features of the star will be imprinted on the planet spectrum, but Doppler shifted based on the radial velocity of the planet. Not including the stellar features in the planet spectrum will therefore underestimate the amount of spectral features at higher spectral resolution.

\subsection{Observatory assumptions}

For the observing strategy, we assume a total integration time of $T_{\mathrm{tot}}=400\,$h per system. In a real survey, the exposure time would not be constant for each star. It would be varied to yield a similar S/N for each planet. However, this is not yet implemented in our simulator and left for future work. Computing the S/N for a fixed exposure time is sufficient given the goals of this analysis.
For PSF subtraction, we assume a pair-wise PSF subtraction such as angular differential imaging with two different roll angles, which effectively doubles the variance of all sources of noise except the photon noise of the planet itself. This should be a conservative noise assumption for the uncorrelated component of the noise. For example, PSF subtraction with a significantly brighter reference star would not necessarily increase the noise of the science images significantly.

For the spectrograph, we center a spectral filter with a 20\% bandwidth (BW) on $\lambda=750\,$nm, which includes the deepest O$_2$ spectral features in the visible. We vary the spectral resolution from $R=20-30,000$. To significantly speed up the computations, we only broaden the model spectra to the resolution of the instrument, but we do not separately integrate for the spectral bin width. This choice implies that our definition of spectral resolution already includes the pixel broadening on the detector, which will differ slightly from the spectral resolution estimated directly from the optical line spread function. Due to the choice of a Gaussian kernel for broadening the spectra, this approach is only acceptable if the spectral bin size is no smaller than the full width at half maximum (FWHM) of the optical line spread function.

The number of detector pixels over which the light of a single spectral bin is spread is given by $N_{\mathrm{pix/bin}}\approx8$. This assumes that the wavelength axis is Nyquist sampled at $750\,$nm ($N_\mathrm{bin/res}=2$) and that the spatial resolution element is Nyquist sampled around $570\,$nm ($0.015^{\prime\prime}$ platescale).

However, Nyquist sampling is not a fundamental requirement for an instrument. The sampling is for example most important to limit systematics when measuring astrometry in the spatial direction, or radial velocities in the spectral direction. However, Nyquist sampling can be relaxed when the detector noise is the primary concern. An example of this is the JWST/NIRSpec IFU \cite{Boker2022}, which is spatially undersampled at all wavelengths ($1-5\,\mu$m). We therefore also run an example with an undersampled wavelength axis ($N_{\mathrm{bin/res}}=1$).

For the detector noise, we first assume a dark current ($\xi=3 \times 10^{-5}$ phot/s)corresponding to a LUVOIR analog\cite{LUVOIR2019arXiv191206219T}, zero read noise, and clock-induced charges (CIC$=2.1 \times 10^{-3}$ phot/$T_{\mathrm{exp}}$). The detector noise will be varied in the following. $T_{\mathrm{exp}}$ is the exposure time for individual images. Unlike the dark current, the read noise and CIC are proportional to the number of individual exposures ($T_{\mathrm{tot}}/T_{\mathrm{exp}}$) instead of the total exposure time ($T_{\mathrm{tot}}$). The CIC dominates the detector noise for individual exposure times $T_{\mathrm{exp}}<70\,$s.
We therefore assume individual exposure times to be $T_\mathrm{exp}=300\,$s, which will be dark current limited.
From there, we explore the dependence of the S/N on the detector noise level by testing a case where these three values (CIC, RN, $\xi$) are divided by a factor ten, and also a bounding case where they are all set to zero.

\begin{figure}
\begin{center}
\begin{tabular}{c}
    \includegraphics[width=0.5\linewidth]{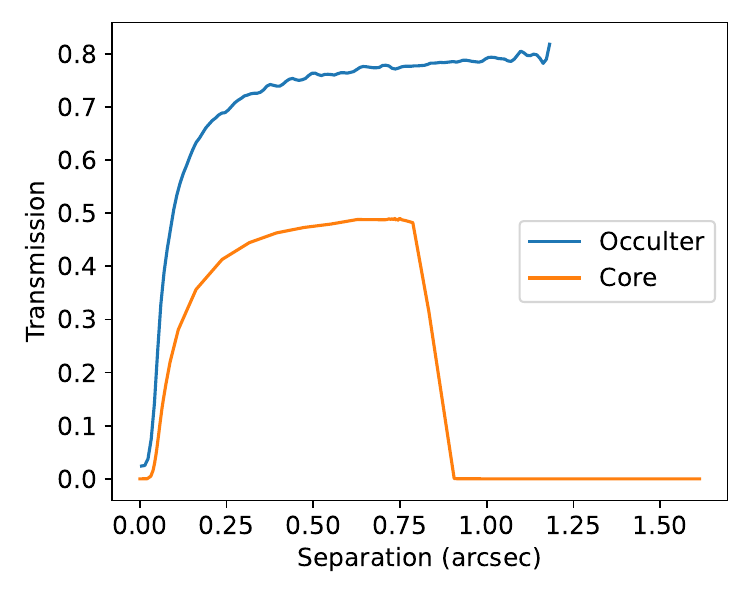}
\end{tabular}
\end{center}
\caption 
{ \label{fig:throughput} Transmission assumptions for the coronagraph. The core throughput ($\tau_{\mathrm{core}}$) is used for planet flux calculations. 
The occulter transmission ($\tau_{\mathrm{occ}}$) is used for the zodi and exozodi.} 
\end{figure}

For the coronagraph, we use the model from the Ultrastable Observatory Roadmap Team (USORT) optical vortex coronagraph (OVC) model as generated by the Coronagraph Design Survey\cite{2024CDS}. We define the separation-dependent starlight suppression as $\zeta$ and we set a starlight suppression floor at $\zeta_{\mathrm{floor}}=10^{-10}$. We will vary $\zeta_{\mathrm{floor}}$ in the following.
 Future work is expected to explore more realistic coronagraph models.
Then, a fraction of the starlight can be subtracted in post processing. We defined previously the ratio of the residual speckles to the original starlight contamination as $g$. The correlated systematic floor after PSF subtraction is therefore set by $g \zeta$. 
For the off-axis transmission of the planet signal, we use the core throughput $\tau_{\mathrm{core}}$ from the same coronagraph model\cite{2024CDS}. Similarly, we use the occulter transmission for extended background sources. Both transmission curves are shown in Fig.~\ref{fig:throughput}.

\subsection{The broadband fluxes}

In the following, we use the names and conventions used in literature where possible, for example \texttt{EXOSIMS} and other similar works\cite{Savransky2017ascl.soft06010S,Nemati2014SPIE.9143E..0QN,Nemati2020JATIS...6c9002N,Stark2019JATIS...5b4009S}. We first define the broadband photon count rates in this section, which are defined with an upper case $C$ for the variable name.

The total unobstructed stellar photon count rate in the spectral bandpass is given by
\begin{equation}
    C_s = F_s\,A\,\tau_\mathrm{opt}\,\mathrm{QE},
\end{equation}
where $F_s$ is the total spectral flux of the star in the observing spectral bandpass in units of phot/m$^2$/s. It is calculated within \texttt{EXOSIMS} for each star in the catalog.
$A$ is the photon collecting area of the observatory and $\tau_\mathrm{opt}$ is the total transmission of the optics. $\mathrm{QE}$ is the quantum efficiency of the detector.

The local starlight contamination in the planet spectrum is therefore given by \cite{Mennesson2024JATIS..10c5004M}:

\begin{equation}
    C_{s,\mathrm{cont}} = C_s\,\zeta \tau_{\mathrm{core}},
\end{equation}

which sets the stellar photon noise at the position of the planet. This implies that the starlight contamination (i.e., raw speckles) is the result of the coronagraph starlight suppression $\zeta$ and off-axis throughput $\tau_{\mathrm{core}}$.

We then measure the level of correlated noise with: 
\begin{equation}
    C_{s,\mathrm{corr}} = C_s\,g\,\zeta \tau_{\mathrm{core}},
\end{equation}
We note that the integrated correlated noise over the bandpass is effectively zero after speckle subtraction. $C_{s,\mathrm{corr}}$ is therefore not the integrated correlated residual speckle photon rate, but it should be interpreted as a proxy to illustrate the importance of the correlated noise compared to other noise sources as seen later in Figure~\ref{fig:noisebudget}. The correlated noise will be defined by its standard deviation per spectral bin in Section~\ref{sec:signoisedeg}.

The magnitude of the zodiacal light per angle area is set to $Z=23\,$mag/arcsec$^{2}$ independent of the pointing direction.
The zodiacal photon count rate is defined as:
\begin{equation}
    C_z = 10^{-0.4\,Z}\,\mathrm{ZP}_V\,\Omega\,A\,\tau_\mathrm{opt}\,\mathrm{QE}\,\tau_{\mathrm{occ}},
\end{equation}
where $\mathrm{ZP}_V$ is the Vega zero point of the spectrograph bandpass. $\Omega=2.26(\lambda/D)^2$ is the area of the aperture (i.e., core) of the off-axis PSF. 

The exozodiacal light is calculated from
\begin{equation}
    C_\mathrm{ez} = \mathrm{JEZ}\,\Omega \,A\,\tau_\mathrm{opt}\,\mathrm{QE}\,\tau_{\mathrm{occ}},
\end{equation}
where $\mathrm{JEZ}$ is the intensity of exo-zodiacal light in units of phot/s/m$^2$/as$^2$, which is itself calculated from
\begin{equation}
    \mathrm{JEZ} = \mathrm{JEZ}_0\, n_\mathrm{ez}\, \left(\frac{1\,\mathrm{au}}{d_\mathrm{eed}}\right)^2.
    \label{eq:JEZ}
\end{equation}
$\mathrm{JEZ}_0$ is the exozodi value at 1 AU for each star in the target list based on the solar system level, which is computed within \texttt{EXOSIMS}. $n_\mathrm{ez}$ is a scaling factor of the exozodi level empirically set to 3 zodis based on the LBTI/Hunt for Observable Signatures of Terrestrial Systems survey for exozodiacal dust\cite{Ertel2020AJ....159..177E}.

The planet light is given by:
\begin{equation}
C_\mathrm{p} = C_s\,p\,\left(\frac{R_p}{d_\mathrm{eed}}\right)^2 \phi\,\tau_{\mathrm{core}}.
\end{equation}
As a reminder, $p$ is the geometric albedo of the planet, and $\phi$ is the phase function.

We define the broadband detector noise photon count rates such that:
\begin{equation}
C_\mathrm{Dark}=\sum_i\,c_\mathrm{Dark,i};\,\,C_\mathrm{RN}=\sum_i\,c_\mathrm{RN,i};\,\,C_\mathrm{CIC}=\sum_i\,c_\mathrm{CIC,i}
\end{equation}
with the count rates per spectral bin $c$ defined in the next section. While the detector noise is better defined for the individual spectral bins, the integrated detector noise over the entire spectral band is useful in the noise budget to compare to the other astrophysical sources of noise.

\subsection{The fluxes per spectral bin}

The template matching S/N requires the photon counts to be known in each individual spectral bin, which we define in this section. We will be using a lower case $c$ for the photon count rates per spectral bin.  

We define the dark current as:

\begin{equation}
c_\mathrm{Dark} = \xi \times N_{\mathrm{pix/bin}}
\end{equation}

where $\xi$ is the dark current photon rate per pixel. The dark current noise is a Poisson noise so $c_\mathrm{Dark}T_\mathrm{tot}$ is its variance.

We assume the read noise ($c_\mathrm{RN}$) to be zero and the effective count rate for the clock-induced charges to be:
\begin{equation}
c_\mathrm{CIC} = \mathrm{CIC}\,N_{\mathrm{pix/bin}}/T_\mathrm{exp}.
\end{equation}
Like for the dark current, $c_\mathrm{CIC}T_\mathrm{tot}$ is the variance if the CIC noise.

In this study, we assume that the exozodi and zodi spectra are uniform across the spectral bandpass of the instrument, such that:
\begin{equation}
c_\mathrm{z,i}=C_\mathrm{z,i}/N_\mathrm{bins};\,\,c_\mathrm{ez,i}=C_\mathrm{ez,i}/N_\mathrm{bins},
\end{equation}
with $N_\mathrm{bins}$ the number of spectral bins in the spectrum. In practice, the exozodi and the zodi should reflect the spectrum of the star and the Sun respectively, but we leave such more detailed simulations for future work. Imperfect modeling of these spectral features could contribute to the correlated noise budget.

For the planet signal, we define $\mathcal{P}(\lambda)$ as the planet albedo spectrum broadened at the resolution of the instrument and $f(\lambda)$ the filter profile of the bandpass. The planet photon count rate per spectral bin is then defined as:
\begin{equation}
    c_{p,i} \propto \mathcal{P}(\lambda_i)f(\lambda_i),\,\mathrm{such\,that}\,C_p = \sum_i c_{p,i},
\end{equation}
where $\lambda_i$ is the central wavelength of each spectral bin.

The residual starlight contamination spectrum follows the same principle:
\begin{equation}
    c_{\mathrm{cont},i} \propto \mathcal{S}(\lambda_i)f(\lambda_i),\,\mathrm{such\,that}\,C_{s,\mathrm{cont}} = \sum_i c_{\mathrm{cont},i},
\end{equation}
with $\mathcal{S}(\lambda)$ the broadened spectrum of the star. This ensures that the speckle intensity per spectral bin integrates to the expected broadband starlight contamination intensity over the assumed spectral bandpass. We do not yet model any continuum modulation of the speckles other than through the covariance matrix.

\subsection{Signal and noise definition}
\label{sec:signoisedeg}

The planet signal, or model, in Equation~\ref{eq:tmsnr} is then simply calculated by scaling from the total integration time $T_\mathrm{tot}$ such that we have 
\begin{equation}
\mathbf{m}=[T_\mathrm{tot}c_{p,1},T_\mathrm{tot}c_{p,2},\dots]^\top 
\end{equation}

Combining all uncorrelated sources of noise, we have:
\begin{equation}
    \sigma_{\mathrm{uncorr},i}^2 = T_\mathrm{tot}c_{p,i} + 2\times T_\mathrm{tot} \left( c_{\mathrm{cont},i} + c_\mathrm{z,i} +c_\mathrm{ez,i} + c_\mathrm{Dark,i} + c_\mathrm{RN,i} + c_\mathrm{CIC,i}\right)
\end{equation}
This accounts for the doubling of the noise resulting from the increased noise from PSF subtraction assuming pair-wise ADI.

For the correlated portion of the noise, we have:
\begin{equation}
    \sigma_{\mathrm{corr},i} = g T_\mathrm{tot} c_{\mathrm{cont},i}
\end{equation}

This implies that the standard deviation of the correlated noise in each bin (i.e., in the speckle subtracted data) is proportional to the speckle intensity of the raw PSF.
The covariance $\Sigma$ is then defined according to Section~\ref{sect:cov}.

\begin{figure}
\begin{center}
\begin{tabular}{c}
    \includegraphics[width=\linewidth]{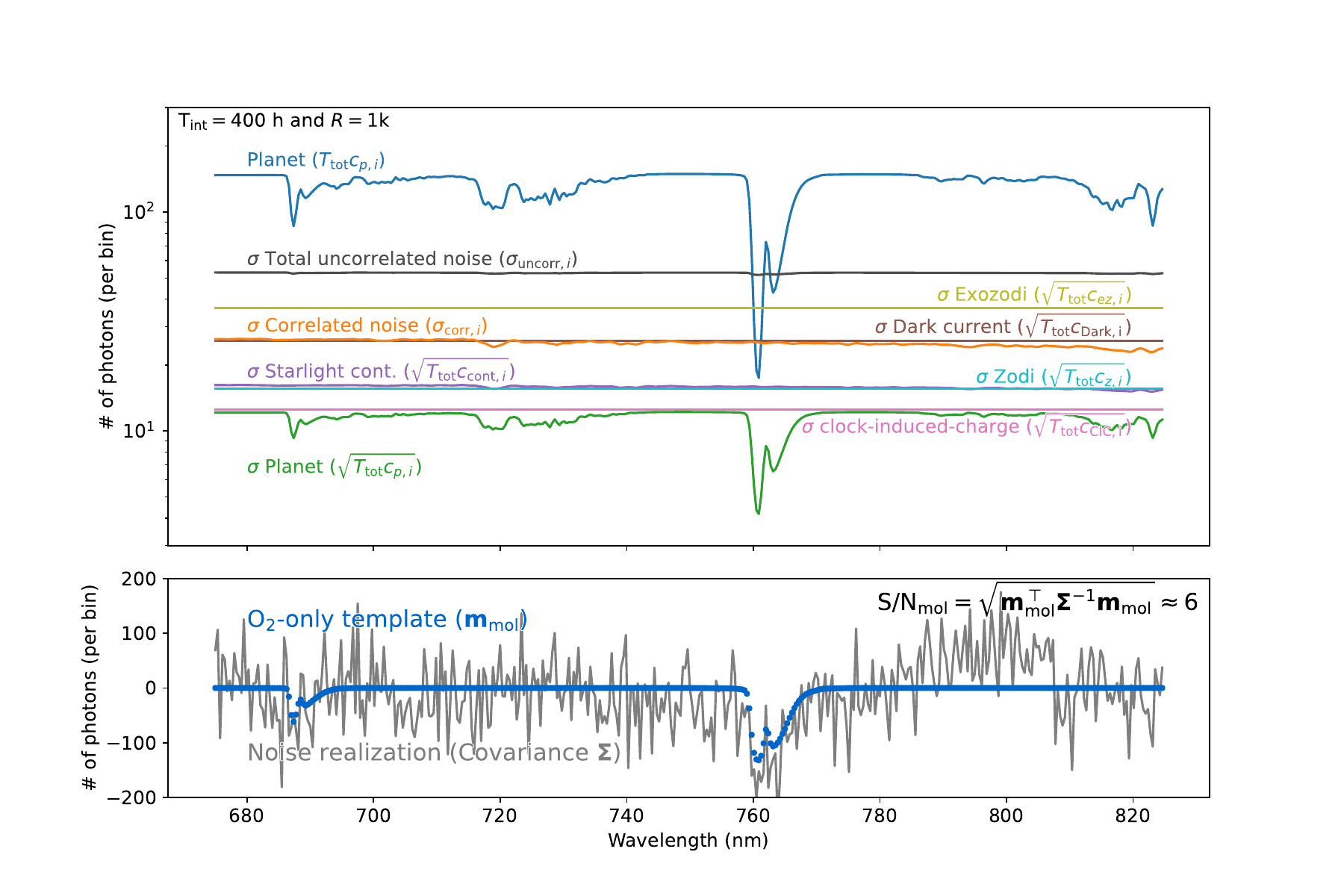}
\end{tabular}
\end{center}
\caption 
{ \label{fig:samplespec} 
Mock observation of an exoEarth around the star HIP 79672 (G2V; $\sim14\,$pc). We assume a 400-hour exposure time and a spectral resolution of $R=1,000$ for a Nyquist-sampled spectrograph. A starlight suppression of $10^{-10}$ is assumed for the coronagraph and the correlated noise floor is set to $10^{-11}$ (i.e., $g=0.1$). The detector noise level is set to a LUVOIR-analog: dark current of $\xi=3 \times 10^{-5}\,$phot/s, zero read noise, and clock-induced charge set to CIC$=2.1 \times 10^{-3}\,$phot/s.
(Top) Total photon count per spectral bin for the planet signal compared to the standard deviation of the different sources of noise.
(Bottom) Oxygen-only template with its envelope subtracted. An example of Gaussian noise realization defined by the covariance matrix $\Sigma$ is provided. The oxygen detection in this example has an expected S/N$\approx6$ (Eq.~\ref{eq:molsnr}).
} 
\end{figure}

The spectra of the planet and various noise terms are illustrated in Fig.~\ref{fig:samplespec} for a $R=1,000$ spectrograph.

\begin{figure}
\begin{center}
\begin{tabular}{c}
    \includegraphics[width=\linewidth]{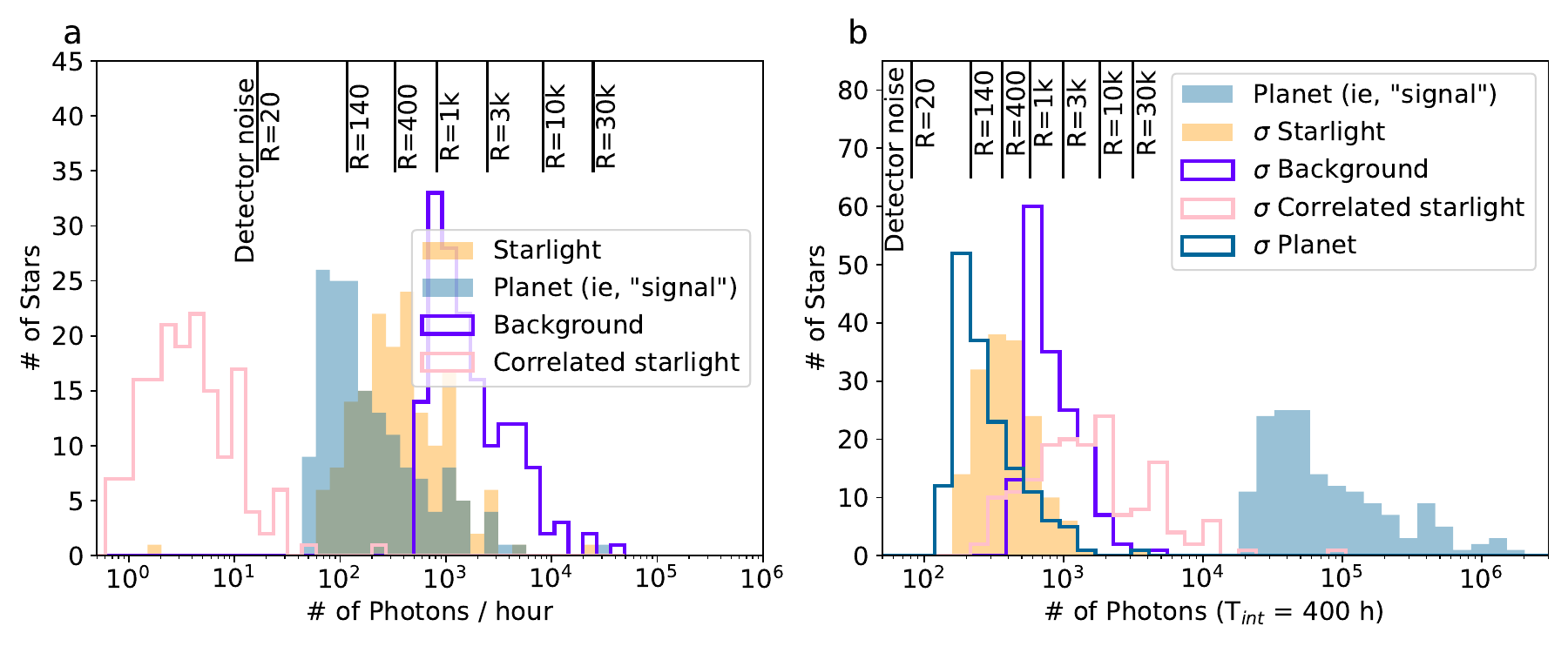}
\end{tabular}
\end{center}
\caption 
{ \label{fig:noisebudget}
Noise budget for a sample of 164 HWO stars. We assume a LUVOIR-analog detector noise level with $10^{-10}$ starlight suppression and $10^{-12}$ stability floor.  (Left) Photon count rates per hour for the planet ($C_\mathrm{p}$) and the different sources of noise, which includes the background light (zodi and  exozodi; $C_\mathrm{z}+C_\mathrm{ez}$), the starlight (uncorrelated $C_{\mathrm{s},\mathrm{cont}}$ or correlated  $C_{\mathrm{s},\mathrm{corr}}$). Each histogram shows the distribution of the photon count rate over the 164 stars of the ExEP HWO star catalog. The astrophysical broadband photon counts do not depend on spectral resolution. Conversely,the summed variance of the detector noise ($C_\mathrm{dark}+C_\mathrm{RN}+C_\mathrm{CIC}$) does not depend on the star, so its histogram would be a delta function. This is why we instead show the detector noise level as a vertical bar at the top of the panel, which is a function of the spectral resolution. This representation is useful to determine the spectral resolution at which the detector noise starts dominating the noise budget.
All the photon count rates other than the planet have been doubled to account for a pair-wise PSF subtraction.
 (Right) Similar to the left panel but showing the standard deviation of the different noise terms given a 400-hour total integration time.} 
\end{figure}

We show an example of the distributions of photon count rates and noise terms over the 164 stars in the target list in Fig.~\ref{fig:noisebudget}. 
This figures illustrates that a $10^{-10}$ coronagraph yield residual starlight at a comparable level as typical exoEarths.
This figure also shows that the detector noise does not dominate the noise budget until moderate spectral resolutions (R$>$1,000) even for detector noise levels of current detector technologies due to the exozodi level. It also highlights how the correlated systematic noise floor dominates the noise budget for long exposure times necessary for exo-Earth characterization. Indeed, only the photon noise is reduced by increasing the exposure time, but the correlated noise standard deviation remains proportional to the starlight contamination. In this case, the speckle subtraction would need to have residual correlated systematics at least $\sim$1000 times fainter than the raw starlight at the position of the planet.

\section{Simulation results}
\label{sect:results}

The goal of this work is to identify some of the main parameters driving the optimal spectral resolution for biosignature searches. We will therefore compute the S/N of a representative population of Earth analogs as a function of detector noise assumptions and the wavefront correlated noise floor of the observations. The code to reproduce these simulations is provided on Github\footnote{\url{https://github.com/jruffio/EXOSIMS_MHRS_scripts}}.

\subsection{Impact of detector noise}

\begin{figure}
\begin{center}
\begin{tabular}{c}
    \includegraphics[width=\linewidth]{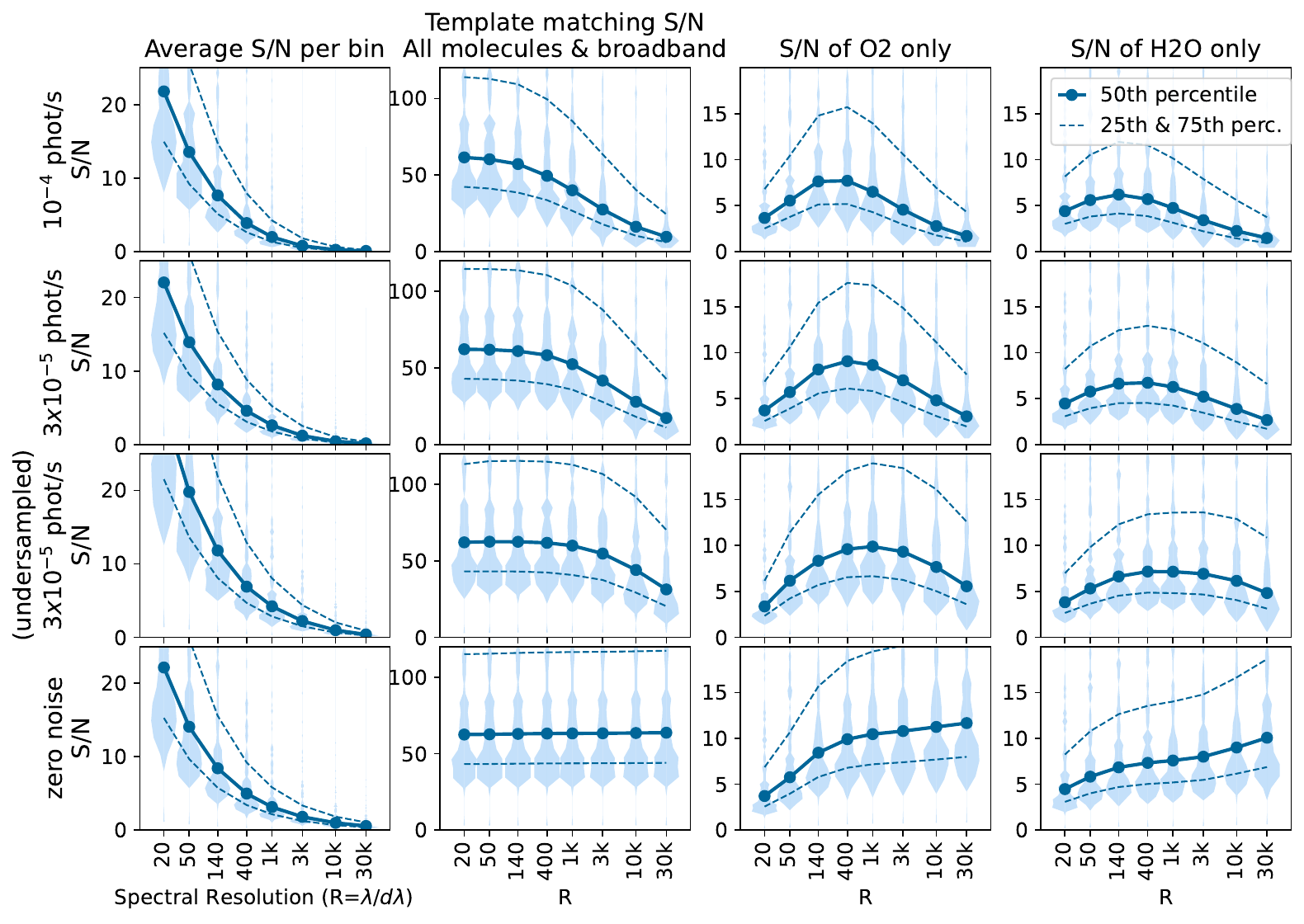}
\end{tabular}
\end{center}
\caption 
{ \label{fig:snr_vs_detec}
The spectral resolution dependence of different definitions of S/N for Earth analogs. Each vertical S/N histogram represents the distribution over the 164 stars in the HWO target list for a fixed set of observatory parameters. In each panel, the 25, 50, and 75 percentiles are drawn. Each row represents one of the four cases varying the detector noise level and the spectral sampling. From left to right, the columns include the S/N per spectral bin, the template matching S/N from Eq.~\ref{eq:tmsnr}, the molecular detection S/N from Eq.~\ref{eq:molsnr} for H$_2$O, and finally the same for O$_2$. In all cases, the stability floor is set to $g\zeta_{\mathrm{floor}}=10^{-13}$.
} 
\end{figure}

First, we fix the correlated noise floor to a negligible level with $g\zeta_{\mathrm{floor}}=10^{-13}$ (i.e., $g=0.001$) and vary the detector noise assumption. We simulate four cases based on the detector noise: 1) $10^{-4}\,$phot/s dark current and $8 \times 10^{-3}$ phot/$T_{\mathrm{exp}}$ CIC corresponding to an order of magnitude improvement compared to the Roman CGI detector\footnote{\url{https://roman.ipac.caltech.edu/page/param-db}} and with Nyquist sampled spectra $N_{\mathrm{bin/res}}=2$, 2) LUVOIR-analog with $3\times10^{-5}\,$phot/s dark current, $2.1 \times 10^{-3}$ phot/$T_{\mathrm{exp}}$ CIC and Nyquist sampled, 3) a similar LUVOIR-analog case, but with undersampled spectra $N_{\mathrm{bin/res}}=1$, and finally 4) a case with zero detector noise. The results of the four sets of simulations are shown in Fig.~\ref{fig:snr_vs_detec}.

We make the following observations:
\begin{itemize}
    \item The S/N per bin decreases with spectral resolution as expected, but it follows neither the broadband template matching S/N nor the molecular detection S/N. In other words, the S/N per bin, or S/N per resolution element, that needs to be achieved to detect biosignatures is a strong function of the spectral resolution.
    \item The broadband template matching S/N, which is relevant for planet detection, only starts decreasing when the detector noise dominates the noise budget. The S/N loss is $\sim16$\% at R$\sim$1000 in the LUVOIR analog case 2). If the detector noise penalty for higher spectral resolutions is negligible, we should also consider the possibility to use the spectrograph for the initial phase of the survey to detect planets instead of a separate imaging mode.
    \item At low spectral resolution, the molecular detection S/N first increases with increasing spectral resolution. This remains true until the detector noise begins to dominate after which the S/N decreases with increasing spectral resolution. The optimal spectral resolution will therefore depend on the detector noise level. \textit{The optimal spectral resolution could be arbitrarily large for a sufficiently low-noise detector}.
\end{itemize}

\subsection{Impact of PSF stability floor}

\begin{figure}
\begin{center}
\begin{tabular}{c}
    \includegraphics[width=\linewidth]{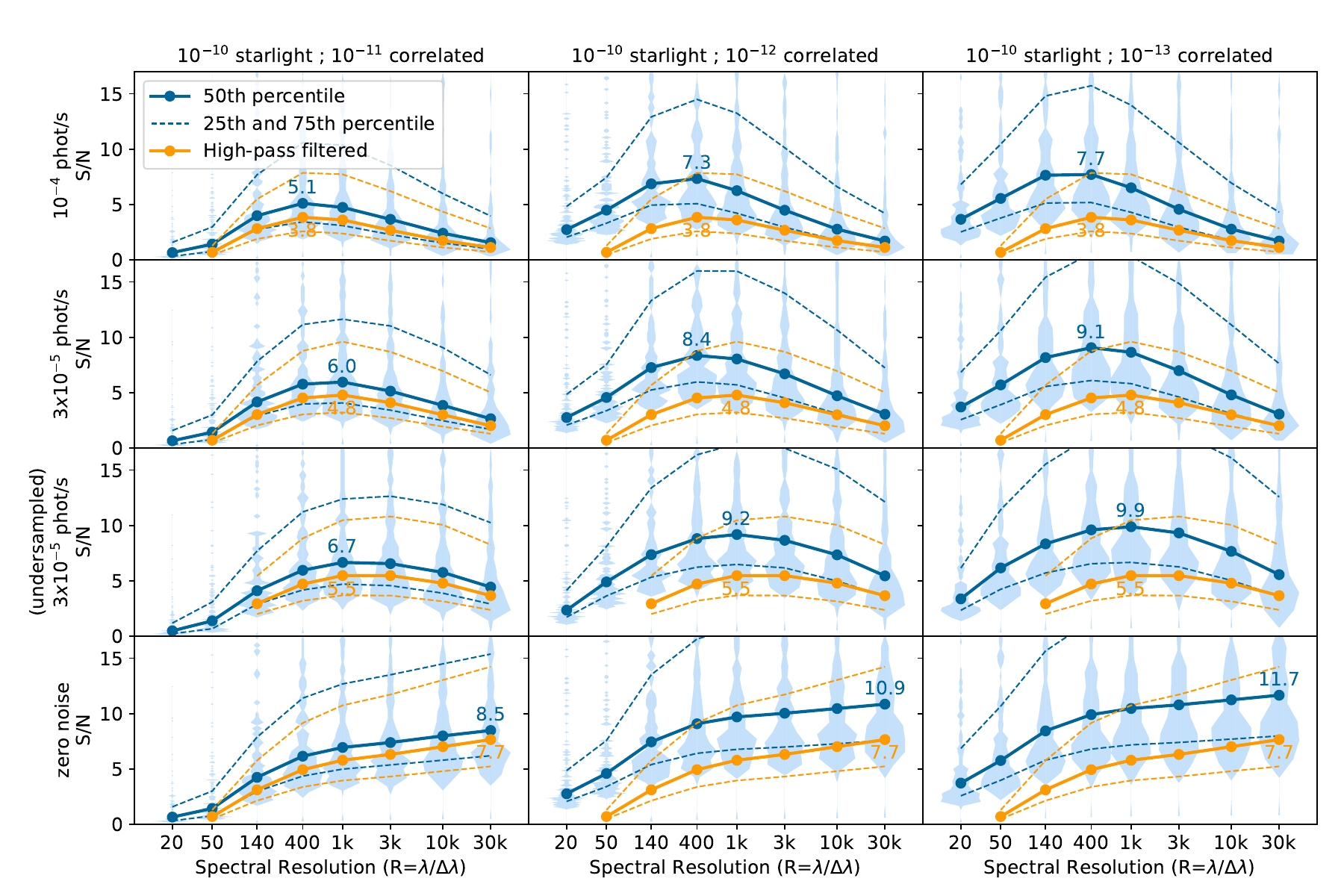}
\end{tabular}
\end{center}
\caption 
{ \label{fig:PSDD_O2}
Impact of the correlated noise floor and detector noise level on the O$_2$ detection S/N. Each sub-panel corresponds to the third column of Fig.~\ref{fig:snr_vs_detec} with the detector noise varying from top to bottom. 
From top to bottom, we vary the level of dark current from $10^{-4}\,$phot/s to a zero-noise detector.
From left to right, the columns corresponds to the correlated noise floor ($g\zeta_{\mathrm{floor}}\in[10^{-11},10^{-12},10^{-13}]$) with the starlight suppression level fixed at $10^{-10}$.
The 25, 50, and 75 percentiles drawn in blue are from Eq.~\ref{eq:molsnr}, while the orange lines are from Eq.~\ref{eq:hpfsnr} with the high-pass filtered S/N. 
} 
\end{figure}

The detector noise level and the correlated noise floor are expected to be the main parameters driving the optimal spectral resolution.
In Fig.~\ref{fig:PSDD_O2}, we therefore show the O$_2$ detection S/N as a function of detector noise, with the same case-studies as in the previous section, and the correlated noise floor ($g\zeta_{\mathrm{floor}}\in[10^{-11},10^{-12},10^{-13}]$). We fix the starlight suppression level at $10^{-10}$ in all cases.
We also plot the detection S/N from Eq.~\ref{eq:hpfsnr}, which corresponds to the case where the spectrum has been high-pass filtered based on the spectral length scale of the correlated noise ($\Lambda_{\mathrm{corr}}=10\,$nm).

From this set of simulations, we make the following observations:
\begin{itemize}
    \item While a single value of the correlation length scale was used, these simulations illustrate a more general consequence of correlated noise: the smaller the correlation length, the higher the minimum required spectral resolution of the spectrograph. Indeed, for a correlated noise floor set at $10^{-11}$, the molecular detection S/N vanishes for spectral resolution that are too close to the resolution corresponding to the correlation scale. For $\Lambda_{\mathrm{corr}}=10\,$nm, the corresponding characteristic spectral resolution for the correlated noise is around $R\approx30$. $R=400$--$1,000$ would provide a substantial improvement in S/N compared to $R=140$ in the case of the LUVOIR analog.
    \item By construction, this high-pass filtered S/N does not depend on the correlated noise floor. It therefore provides a lower limit on the molecule S/N in case of degraded stability performance.
    \item We also note that correlation length and amplitude sets an upper limit on the broadband planet detection. For a $10^{-11}$ stability floor and based on the set of assumptions presented in this work, the maximum median achievable broadband template matching S/N across all spectral resolutions would be at best $\sim$13 with a LUVOIR-analog assumption instead of $\sim$63 with a $10^{-13}$ stability floor.
\end{itemize}

\subsection{Impact of the coronagraph starlight suppression}

\begin{figure}
\begin{center}
\begin{tabular}{c}
    \includegraphics[width=\linewidth]{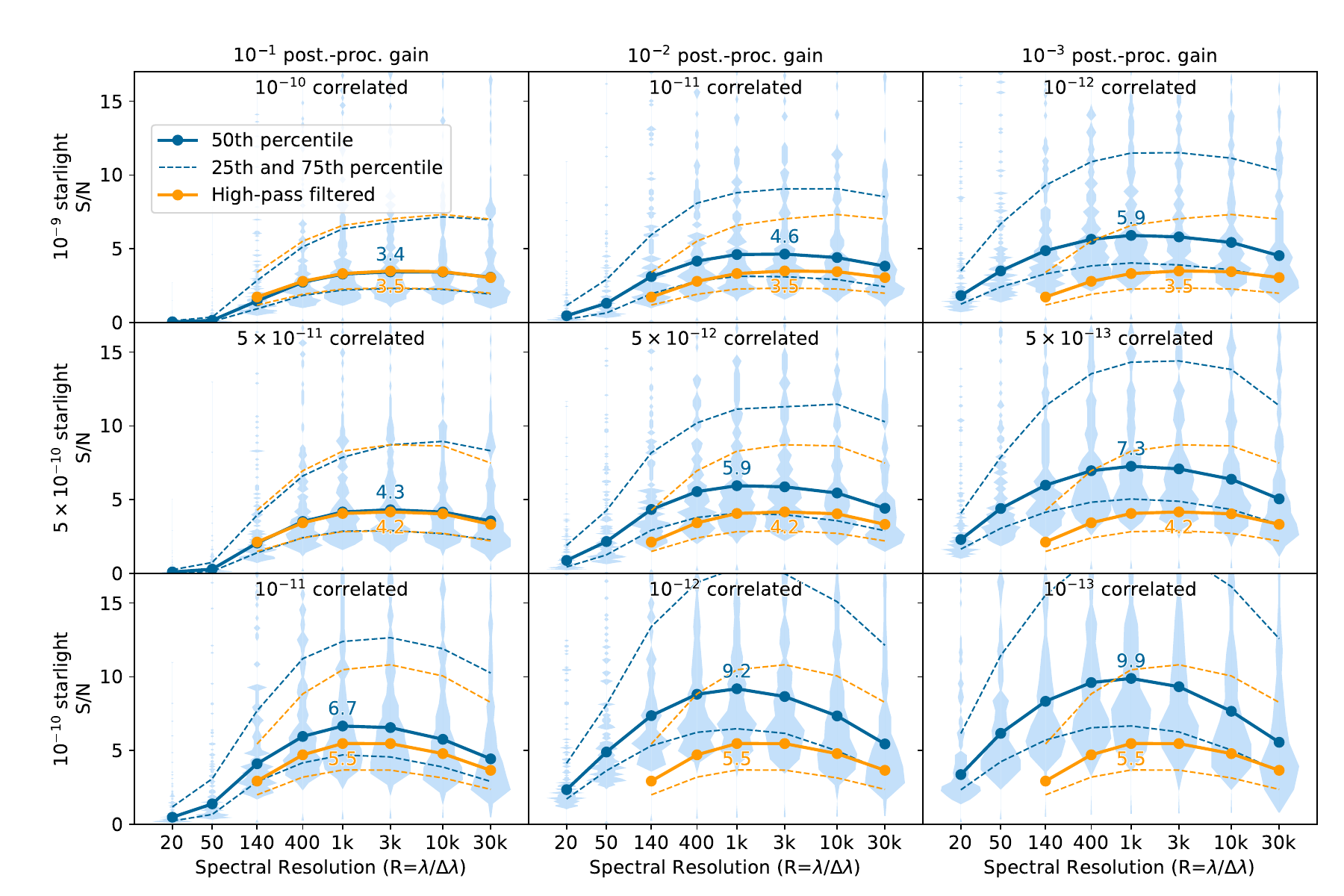}
\end{tabular}
\end{center}
\caption 
{ \label{fig:PSDD_starlight_O2}
Impact of the starlight suppression and post-processing gain on the observatory performance and spectral resolution. Each panel is similar to Fig.~\ref{fig:PSDD_O2}. Each row corresponds to a different level of coronagraph starlight suppression from $10^{-9}$ (top) to $10^{-10}$ (bottom). The post-processing gain varies from 0.1 (left column) to 0.001 (right column). The combination of a starlight suppression and post-processing gain results in different levels of correlated noise, which are indicated in each panel.
} 
\end{figure}

In Figure~\ref{fig:PSDD_starlight_O2}, we aim to illustrate the interplay between the coronagraph starlight suppression and the observatory stability requirements parametrized as the post processing gain. We select the undersampled LUVOIR analog with $3\times10^{-5}\,$phot/s dark current and  $N_{\mathrm{bin/res}}=1$ from the previous section.
We vary the level of starlight suppression from  $10^{-9}$ to $10^{-10}$ and the post-processing gain from $10^{-1}$ to $10^{-3}$.

In this case, we note that:
\begin{itemize}
    \item Improvements of the starlight suppression or the post processing gain improve the overall detection sensitivity.
    \item There is a trade-off between the coronagraph design and the observatory stability requirements. Indeed, Figure~\ref{fig:PSDD_starlight_O2} shows that a $10^{-9}$ coronagraph stable at $10^{-12}$ yields comparable oxygen detection as a $10^{-10}$ coronagraph stable at $10^{-11}$. This same relationship was also seen by [\citenum{SteigerChen2026}] using an S/N per bin formalism for determining a successful detection/characterization. 
    \item A spectral resolution $R\sim1,000$ appears optimal in all cases given the assumed detector noise.
\end{itemize}

\section{Discussion}
\label{sect:discussion}

\subsection{Spectrograph and coronagraph design}
\begin{figure}
\begin{center}
\begin{tabular}{c}
    \includegraphics[width=0.75\linewidth]{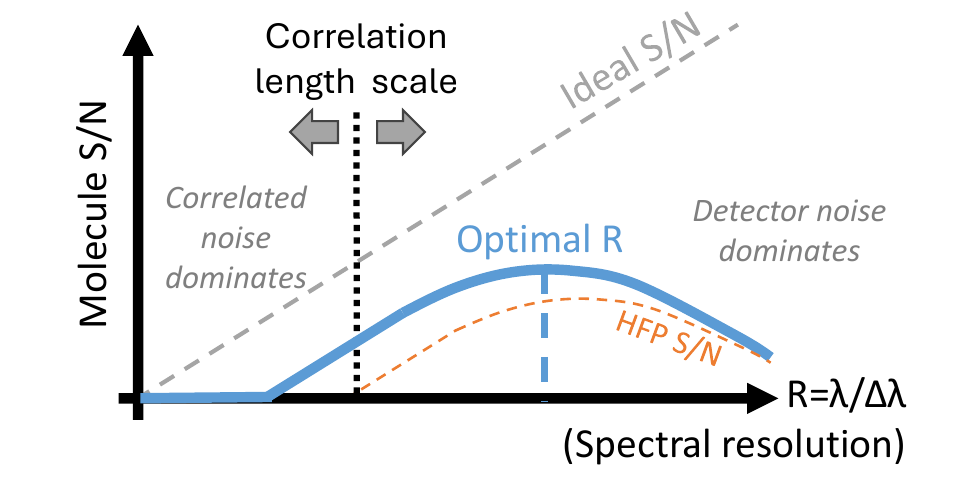}
\end{tabular}
\end{center}
\caption 
{ \label{fig:cartoon}
Conceptual illustration of the spectral resolution optimization. The correlated speckle noise is expected to dominate at low spectral resolutions while detector noise dominates at high spectral resolution. If the amplitude of the correlated noise is significant, its characteristic spectral resolution from its length scale sets a limit on the spectral resolution below wish the molecular S/N vanishes. The high-pass filter (HPF) S/N is the detection significance calculated after subtracting any spectral features with characteristic scales larger than the correlated noise. The HPF S/N is unaffected by the amplitude of the correlated noise and therefore represents the lower limit on the detection S/N.} 
\end{figure}

The goal of this work was to define a framework to optimize the spectral resolution of HWO and explore the effects of the main design parameters. 
Despite the limited scope of this analysis, we are already able to draw some initial conclusions. The first one is the importance of modeling correlated noise with a covariance model in exposure time or yield calculations.
We find that it can have a major impact on the optimal spectral resolution of the observatory. Even $10^{-11}$ correlated noise for a $10^{-10}$ starlight suppression leads to vanishing molecular detection S/N for O$_2$ at low spectral resolutions. In this case, the correlation scale sets a minimum spectral resolution for the instrument. Therefore, depending on the correlated noise floor in the spectra, a moderate to high resolution spectrograph might not only be more sensitive, but it could be necessary for biosignature detection. 
Within our framework, we find that a moderate resolution spectrograph $R>1,000$ would provide better sensitivity to O$_2$ for a LUVOIR-analog detector noise assumption and the correlated noise floor. 
The possibility of zero-noise detectors could even make a compelling case for spectral resolution higher than $R>10,000$.
We summarize the general effects of the correlation length scale, correlation amplitude, and the detector noise on the optimal spectral resolution in Figure~\ref{fig:cartoon}.

\subsection{The correlated noise}

As discussed in the introduction, the sources of correlated noise are diverse and can have an impact at any spectral resolution: PSF chromaticity, fringing, post processing and atmosphere model systematics. However, PSF subtraction residuals (i.e., speckles) are expected to dominate at low spectral resolutions. As an example of systematics, rotating the observatory for ADI observations would lead to a different angle of incidence of the sunlight, which would produce repeatable wavefront errors that are challenging to model and subtract. 
This remains true with forward modeling frameworks, without separate PSF subtraction, in which the planet and the starlight are jointly modeled\cite{Flasseur2018A&A...618A.138F,Ruffio2024AJ....168...73R}. 
We showed how increasing the spectral resolution of the spectrograph can also mitigate the impact of the correlated noise. 
Although the simple covariance model used in this work is useful to highlight the relevant design trade parameters, more accurate models of the correlated noise sources are essential. Work incorporating wavefront stability into these exposure time calculators in order to look at these trades is underway \cite{SteigerChen2026}. It will be critical to update the covariance model with future detailed models of the observatory behavior to estimate accurate molecular S/N and planet yield for the mission. The correlation length of $10\,\mathrm{nm}$ chosen in this work is only slightly larger than the width of the oxygen absorption band (see Figure~\ref{fig:spectra}). Smaller correlation lengths would likely impact the oxygen detection even more, while larger scales would have a diminishing to negligible impact. How different correlation lengths impact the detection of molecules will be a function of the specific spectral signature of the considered molecules. While we have modeled a single correlation length, real correlated noise will feature a distribution of length scales with varying magnitude; this means that the impact of any correlated noise source should not be solely focused on its dominant length scale.

We highlighted that wavefront stability might be more important than raw starlight suppression as the former does not decrease with the square root of the exposure time (Figure~\ref{fig:noisebudget}). 
However, the correlated noise model used in this work does not account for the temporal variations of the correlated noise. This means that the correlated noise floor does not improve with exposure time, which is not accurate in practice. The wavefront errors will have a temporal timescale over which the speckle noise evolves. When dominated by the correlated noise, the S/N will still improve but only as $\sqrt{T_{\mathrm{tot}}/t_\mathrm{corr}}$ with $t_\mathrm{corr}$ defining the timescale of the evolution. This is compared to $\sqrt{T_{\mathrm{tot}}}$ if the S/N is dominated by uncorrelated noise.

\subsection{Concept of operation and post processing (COPP)}

The broadband template matching S/N represents the detection significance of the planet. As long as the correlated noise does not dominate, this S/N is not a strong function of spectral resolution until the detector noise dominates for Earth analogs. There is no penalty in using higher spectral resolution, as opposed to an imager, for planet detection with HWO from a pure photon and detector noise level consideration. 
Therefore, we suggest exploring further the possibility to use the spectrograph for the initial phase of the mission which is to detect and vet planet candidates. Spectroscopic observation would provide additional information that would be valuable for candidate vetting and reducing false positive rates\cite{Agrawal2023AJ....166...15A}. For example, the spectral shape could help distinguish planetary and background sources. With an integral field spectrograph, such a strategy would also provide the spectra of any additional planets in the system ``for free''\cite{Howe2024JATIS..10b5008H}. 
However, unless the correlated noise dominates, MHRS does not necessarily significantly improve the detection sensitivity of Earth-like planets (Fig.~\ref{fig:snr_vs_detec}) because the continuum component of the spectrum dominates the S/N in this case. This highlights the importance of recovering the continuum of the planet even at higher spectral resolutions. Other than PSF variability, another challenge to recover the continuum is the data reduction and spectral extraction of integral field spectrographs. This is currently the main limitation for JWST/NIRSpec\cite{Ruffio2024AJ....168...73R}. The data reduction is made more challenging with spatially or spectrally under-sampled spectrographs, but it is the topic of on-going work and it is not a fundamental limit of the data. Undersampling the spectrograph is useful to reduce the importance of detector noise, when the latter is the primary concern. JWST/NIRSpec is for example spatially undersampled at all wavelengths ($1-5\,\mu$m) with 100\,mas spaxels. 

We note that spatially undersampling the spectrograph could come with additional constraints on concept of operations, for example to ensure that deep observations of a planet occur when the signal is centered on a pixel to avoid unnecessarily splitting the flux over additional pixels. Spatially undersampling the spectrograph could also make the detection of planets more challenging at small project separations. These trade-offs should be explored in more detail in future studies.

\subsection{Gaussian vs. photon noise statistics}

The template matching framework presented in this work relies on the assumption of Gaussian noise and Gaussian likelihood. It is therefore warranted to check the validity of this assumption given the relatively low photon count rates involved in the detection of Earth analogs in reflected light.
A 20\% bandpass with $R\sim1,000$ would lead to 200 spectral bins across the full bandpass assuming one bin per spectral resolution element.
Around a typical HWO star, this would lead to about $~1$~phot/h/bin for the planet and of the order of $~10$~phot/h/bin for each one of the background, starlight and detector noise components (Fig.~\ref{fig:noisebudget}).
Photon noise can be approximated by Gaussian noise when the average photon count is greater than $\sim$20, which would be true for hour-long observations. While the planet photon count is an order of magnitude lower, the relevant quantity is the total photon count from which the noise is calculated.
Finally, the characterization of exo-Earth atmospheres requires hundreds of hours independent of the spectral resolution, which would imply tens of photons per spectral bin even at $R\sim10,000$.

\section{Conclusion}
\label{sect:conclusion}

In this work, we inform the choice of spectral resolution for the spectrograph in the context of HWO. 
We showed that ``template matching'' provides a convenient framework to define the detection S/N of planets or molecules. It is valid at any spectral resolution and rigorously allows the modeling of speckle noise correlations.
However, although related, the proposed statistical framework is limited to molecular detection S/N and does not inform the achieved uncertainties on atmospheric abundances (Wolff et al.; in prep). 
Based on preliminary simulations, we find that a moderate to high-resolution spectrograph ($R>1,000$) is likely more sensitive, and could even prove to be necessary, for the detection of biosignatures depending on the correlated noise floor set by the wavefront stability of the observatory. 
The combination of ultra-low noise detectors and high-resolution spectroscopy, which is less sensitive to correlated residual noise, could provide an opportunity to relax observatory stability requirements.
 This highlights the need for atmospheric models at moderate to high spectral resolutions. 
Increasing the spectral resolution of the spectrograph would also provide new science opportunities for the observatory.
We conclude that it is essential to model the noise covariance in future exposure time calculators and yield simulations for HWO.
The choice of spectral resolution is likely to have major impact on the design of the coronagraph system and spectrograph.
A more exhaustive exploration of the trade-space of observatory design parameters and accurate models of the observatory wavefront variability is necessary before definitively concluding what the optimal spectral resolution is for Earth-like planet detection.

\subsection*{Disclosures}
Minor use of ChatGPT was made to assist in coding, figures, and writing latex equations.

\subsection* {Code, Data, and Materials Availability} 
An example Python script is provided on Github\footnote{\url{https://github.com/jruffio/speckles_snr_gain}}.

\subsection* {Acknowledgments}
 Material presented in this work is supported by the National Aeronautics and Space Administration under Grants/Contracts/Agreements No. 80NSSC25K7300 (B.D., J.-B.R.) issued through the Astrophysics Division of the Science Mission Directorate. Any opinions, findings, and conclusions or recommendations expressed in this work are those of the author(s) and do not necessarily reflect the views of the National Aeronautics and Space Administration.
The research at the Jet Propulsion Laboratory (JPL), California Institute of Technology, was performed under contract with the National Aeronautics and Space Administration. S.S. acknowledges support from an STScI Postdoctoral Fellowship.


\bibliography{report}   

@ARTICLE{Lupu2016AJ....152..217L,
       author = {{Lupu}, Roxana E. and {Marley}, Mark S. and {Lewis}, Nikole and {Line}, Michael and {Traub}, Wesley A. and {Zahnle}, Kevin},
        title = "{Developing Atmospheric Retrieval Methods for Direct Imaging Spectroscopy of Gas Giants in Reflected Light. I. Methane Abundances and Basic Cloud Properties}",
      journal = {\aj},
         year = 2016,
        month = dec,
       volume = {152},
       number = {6},
          eid = {217},
        pages = {217},
          doi = {10.3847/0004-6256/152/6/217},
archivePrefix = {arXiv},
       eprint = {1604.05370},
 primaryClass = {astro-ph.IM},
       adsurl = {https://ui.adsabs.harvard.edu/abs/2016AJ....152..217L}
}

@ARTICLE{Meadows1996JGR...101.4595M,
       author = {{Meadows}, V.~S. and {Crisp}, D.},
        title = "{Ground-based near-infrared observations of the Venus nightside: The thermal structure and water abundance near the surface}",
      journal = {\jgr},
         year = 1996,
        month = jan,
       volume = {101},
       number = {E2},
        pages = {4595-4622},
          doi = {10.1029/95JE03567},
       adsurl = {https://ui.adsabs.harvard.edu/abs/1996JGR...101.4595M}
}

@ARTICLE{Robinson2011AsBio..11..393R,
       author = {{Robinson}, Tyler D. and {Meadows}, Victoria S. and {Crisp}, David and {Deming}, Drake and {A'Hearn}, Michael F. and {Charbonneau}, David and {Livengood}, Timothy A. and {Seager}, Sara and {Barry}, Richard K. and {Hearty}, Thomas and {Hewagama}, Tilak and {Lisse}, Carey M. and {McFadden}, Lucy A. and {Wellnitz}, Dennis D.},
        title = "{Earth as an Extrasolar Planet: Earth Model Validation Using EPOXI Earth Observations}",
      journal = {Astrobiology},
         year = 2011,
        month = jun,
       volume = {11},
       number = {5},
        pages = {393-408},
          doi = {10.1089/ast.2011.0642},
       adsurl = {https://ui.adsabs.harvard.edu/abs/2011AsBio..11..393R}
}

@ARTICLE{Gordon2022JQSRT.27707949G,
       author = {{Gordon}, I.~E. and {Rothman}, L.~S. and {Hargreaves}, R.~J. and {Hashemi}, R. and {Karlovets}, E.~V. and {Skinner}, F.~M. and {Conway}, E.~K. and {Hill}, C. and {Kochanov}, R.~V. and {Tan}, Y. and {Wcis{\l}o}, P. and {Finenko}, A.~A. and {Nelson}, K. and {Bernath}, P.~F. and {Birk}, M. and {Boudon}, V. and {Campargue}, A. and {Chance}, K.~V. and {Coustenis}, A. and {Drouin}, B.~J. and {Flaud}, J.-M. and {Gamache}, R.~R. and {Hodges}, J.~T. and {Jacquemart}, D. and {Mlawer}, E.~J. and {Nikitin}, A.~V. and {Perevalov}, V.~I. and {Rotger}, M. and {Tennyson}, J. and {Toon}, G.~C. and {Tran}, H. and {Tyuterev}, V.~G. and {Adkins}, E.~M. and {Baker}, A. and {Barbe}, A. and {Can{\`e}}, E. and {Cs{\'a}sz{\'a}r}, A.~G. and {Dudaryonok}, A. and {Egorov}, O. and {Fleisher}, A.~J. and {Fleurbaey}, H. and {Foltynowicz}, A. and {Furtenbacher}, T. and {Harrison}, J.~J. and {Hartmann}, J.-M. and {Horneman}, V.-M. and {Huang}, X. and {Karman}, T. and {Karns}, J. and {Kassi}, S. and {Kleiner}, I. and {Kofman}, V. and {Kwabia-Tchana}, F. and {Lavrentieva}, N.~N. and {Lee}, T.~J. and {Long}, D.~A. and {Lukashevskaya}, A.~A. and {Lyulin}, O.~M. and {Makhnev}, V. Yu. and {Matt}, W. and {Massie}, S.~T. and {Melosso}, M. and {Mikhailenko}, S.~N. and {Mondelain}, D. and {M{\"u}ller}, H.~S.~P. and {Naumenko}, O.~V. and {Perrin}, A. and {Polyansky}, O.~L. and {Raddaoui}, E. and {Raston}, P.~L. and {Reed}, Z.~D. and {Rey}, M. and {Richard}, C. and {T{\'o}bi{\'a}s}, R. and {Sadiek}, I. and {Schwenke}, D.~W. and {Starikova}, E. and {Sung}, K. and {Tamassia}, F. and {Tashkun}, S.~A. and {Vander Auwera}, J. and {Vasilenko}, I.~A. and {Vigasin}, A.~A. and {Villanueva}, G.~L. and {Vispoel}, B. and {Wagner}, G. and {Yachmenev}, A. and {Yurchenko}, S.~N.},
        title = "{The HITRAN2020 molecular spectroscopic database}",
      journal = {\jqsrt},
         year = 2022,
        month = jan,
       volume = {277},
          eid = {107949},
        pages = {107949},
          doi = {10.1016/j.jqsrt.2021.107949},
       adsurl = {https://ui.adsabs.harvard.edu/abs/2022JQSRT.27707949G}
}

@ARTICLE{Ruffio2024AJ....168...73R,
       author = {{Ruffio}, Jean-Baptiste and {Perrin}, Marshall D. and {Hoch}, Kielan K.~W. and {Kammerer}, Jens and {Konopacky}, Quinn M. and {Pueyo}, Laurent and {Madurowicz}, Alex and {Rickman}, Emily and {Theissen}, Christopher A. and {Agrawal}, Shubh and {Greenbaum}, Alexandra Z. and {Miles}, Brittany E. and {Barman}, Travis S. and {Balmer}, William O. and {Llop-Sayson}, Jorge and {Girard}, Julien H. and {Rebollido}, Isabel and {Soummer}, R{\'e}mi and {Allen}, Natalie H. and {Anderson}, Jay and {Beichman}, Charles A. and {Bellini}, Andrea and {Bryden}, Geoffrey and {Espinoza}, N{\'e}stor and {Glidden}, Ana and {Huang}, Jingcheng and {Lewis}, Nikole K. and {Libralato}, Mattia and {Louie}, Dana R. and {Sohn}, Sangmo Tony and {Seager}, Sara and {van der Marel}, Roeland P. and {Wakeford}, Hannah R. and {Watkins}, Laura L. and {Ygouf}, Marie and {Mountain}, C. Matt},
        title = "{JWST-TST High Contrast: Achieving Direct Spectroscopy of Faint Substellar Companions Next to Bright Stars with the NIRSpec Integral Field Unit}",
      journal = {\aj},
         year = 2024,
        month = aug,
       volume = {168},
       number = {2},
          eid = {73},
        pages = {73},
          doi = {10.3847/1538-3881/ad5281},
archivePrefix = {arXiv},
       eprint = {2310.09902},
 primaryClass = {astro-ph.EP},
       adsurl = {https://ui.adsabs.harvard.edu/abs/2024AJ....168...73R}
}

@ARTICLE{Stark2019JATIS...5b4009S,
       author = {{Stark}, Christopher C. and {Belikov}, Rus and {Bolcar}, Matthew R. and {Cady}, Eric and {Crill}, Brendan P. and {Ertel}, Steve and {Groff}, Tyler and {Hildebrandt}, Sergi and {Krist}, John and {Lisman}, P. Douglas and {Mazoyer}, Johan and {Mennesson}, Bertrand and {Nemati}, Bijan and {Pueyo}, Laurent and {Rauscher}, Bernard J. and {Riggs}, A.~J. and {Ruane}, Garreth and {Shaklan}, Stuart B. and {Sirbu}, Dan and {Soummer}, Remi and {Laurent}, Kathryn St. and {Zimmerman}, Neil},
        title = "{ExoEarth yield landscape for future direct imaging space telescopes}",
      journal = {Journal of Astronomical Telescopes, Instruments, and Systems},
         year = 2019,
        month = apr,
       volume = {5},
          eid = {024009},
        pages = {024009},
          doi = {10.1117/1.JATIS.5.2.024009},
archivePrefix = {arXiv},
       eprint = {1904.11988},
 primaryClass = {astro-ph.EP},
       adsurl = {https://ui.adsabs.harvard.edu/abs/2019JATIS...5b4009S}
}

@ARTICLE{Nemati2020JATIS...6c9002N,
       author = {{Nemati}, Bijan and {Stahl}, H. Philip and {Stahl}, Mark T. and {Ruane}, Garreth J. and {Sheldon}, Leah J.},
        title = "{Method for deriving optical telescope performance specifications for Earth-detecting coronagraphs}",
      journal = {Journal of Astronomical Telescopes, Instruments, and Systems},
         year = 2020,
        month = jul,
       volume = {6},
          eid = {039002},
        pages = {039002},
          doi = {10.1117/1.JATIS.6.3.039002},
       adsurl = {https://ui.adsabs.harvard.edu/abs/2020JATIS...6c9002N}
}

@ARTICLE{Mennesson2024JATIS..10c5004M,
       author = {{Mennesson}, Bertrand and {Belikov}, Ruslan and {Por}, Emiel and {Serabyn}, Eugene and {Ruane}, Garreth and {Riggs}, A.~J. Eldorado and {Sirbu}, Dan and {Pueyo}, Laurent and {Soummer}, Remi and {Kasdin}, Jeremy and {Shaklan}, Stuart and {Seo}, Byoung-Joon and {Stark}, Christopher and {Cady}, Eric and {Chen}, Pin and {Crill}, Brendan and {Fogarty}, Kevin and {Greenbaum}, Alexandra and {Guyon}, Olivier and {Juanola-Parramon}, Roser and {Kern}, Brian and {Krist}, John and {Macintosh}, Bruce and {Marx}, David and {Mawet}, Dimitri and {Prada}, Camilo Mejia and {Morgan}, Rhonda and {Nemati}, Bijan and {Pogorelyuk}, Leonid and {Redmond}, Susan and {Seager}, Sara and {Siegler}, Nicholas and {Stapelfeldt}, Karl and {Steiger}, Sarah and {Trauger}, John and {Wallace}, James K. and {Ygouf}, Marie and {Zimmerman}, Neil},
        title = "{Current laboratory performance of starlight suppression systems and potential pathways to desired Habitable Worlds Observatory exoplanet science capabilities}",
      journal = {Journal of Astronomical Telescopes, Instruments, and Systems},
         year = 2024,
        month = jul,
       volume = {10},
          eid = {035004},
        pages = {035004},
          doi = {10.1117/1.JATIS.10.3.035004},
archivePrefix = {arXiv},
       eprint = {2404.18036},
 primaryClass = {astro-ph.IM},
       adsurl = {https://ui.adsabs.harvard.edu/abs/2024JATIS..10c5004M}
}

@ARTICLE{Feng2018AJ....155..200F,
       author = {{Feng}, Y. Katherina and {Robinson}, Tyler D. and {Fortney}, Jonathan J. and {Lupu}, Roxana E. and {Marley}, Mark S. and {Lewis}, Nikole K. and {Macintosh}, Bruce and {Line}, Michael R.},
        title = "{Characterizing Earth Analogs in Reflected Light: Atmospheric Retrieval Studies for Future Space Telescopes}",
      journal = {\aj},
         year = 2018,
        month = may,
       volume = {155},
       number = {5},
          eid = {200},
        pages = {200},
          doi = {10.3847/1538-3881/aab95c},
archivePrefix = {arXiv},
       eprint = {1803.06403},
 primaryClass = {astro-ph.EP},
       adsurl = {https://ui.adsabs.harvard.edu/abs/2018AJ....155..200F}
}

@ARTICLE{Brandt2014PNAS..11113278B,
       author = {{Brandt}, T.~D. and {Spiegel}, D.~S.},
        title = "{Prospects for detecting oxygen, water, and chlorophyll on an exo-Earth}",
      journal = {Proceedings of the National Academy of Science},
         year = 2014,
        month = sep,
       volume = {111},
       number = {37},
        pages = {13278-13283},
          doi = {10.1073/pnas.1407296111},
archivePrefix = {arXiv},
       eprint = {1404.5337},
 primaryClass = {astro-ph.EP},
       adsurl = {https://ui.adsabs.harvard.edu/abs/2014PNAS..11113278B}
}

@ARTICLE{Greco2016ApJ...833..134G,
       author = {{Greco}, Johnny P. and {Brandt}, Timothy D.},
        title = "{The Measurement, Treatment, and Impact of Spectral Covariance and Bayesian Priors in Integral-field Spectroscopy of Exoplanets}",
      journal = {\apj},
         year = 2016,
        month = dec,
       volume = {833},
       number = {2},
          eid = {134},
        pages = {134},
          doi = {10.3847/1538-4357/833/2/134},
archivePrefix = {arXiv},
       eprint = {1602.00691},
 primaryClass = {astro-ph.EP},
       adsurl = {https://ui.adsabs.harvard.edu/abs/2016ApJ...833..134G}
}

@ARTICLE{Mamajek2024arXiv240212414M,
       author = {{Mamajek}, Eric and {Stapelfeldt}, Karl},
        title = "{NASA Exoplanet Exploration Program (ExEP) Mission Star List for the Habitable Worlds Observatory (2023)}",
      journal = {arXiv e-prints},
         year = 2024,
        month = feb,
          eid = {arXiv:2402.12414},
        pages = {arXiv:2402.12414},
          doi = {10.48550/arXiv.2402.12414},
archivePrefix = {arXiv},
       eprint = {2402.12414},
 primaryClass = {astro-ph.IM},
       adsurl = {https://ui.adsabs.harvard.edu/abs/2024arXiv240212414M}
}

@ARTICLE{Savransky2010PASP..122..401S,
       author = {{Savransky}, Dmitry and {Kasdin}, N. Jeremy and {Cady}, Eric},
        title = "{Analyzing the Designs of Planet-Finding Missions}",
      journal = {\pasp},
         year = 2010,
        month = apr,
       volume = {122},
       number = {890},
        pages = {401},
          doi = {10.1086/652181},
archivePrefix = {arXiv},
       eprint = {0903.4915},
 primaryClass = {astro-ph.IM},
       adsurl = {https://ui.adsabs.harvard.edu/abs/2010PASP..122..401S}
}

@software{Savransky2017ascl.soft06010S,
       author = {{Savransky}, Dmitry and {Delacroix}, Christian and {Garrett}, Daniel},
        title = "{EXOSIMS: Exoplanet Open-Source Imaging Mission Simulator}",
 howpublished = {Astrophysics Source Code Library, record ascl:1706.010},
         year = 2017,
        month = jun,
          eid = {ascl:1706.010},
       adsurl = {https://ui.adsabs.harvard.edu/abs/2017ascl.soft06010S}
}

@ARTICLE{Savransky2016JATIS...2a1006S,
       author = {{Savransky}, Dmitry and {Garrett}, Daniel},
        title = "{WFIRST-AFTA coronagraph science yield modeling with EXOSIMS}",
      journal = {Journal of Astronomical Telescopes, Instruments, and Systems},
         year = 2016,
        month = jan,
       volume = {2},
          eid = {011006},
        pages = {011006},
          doi = {10.1117/1.JATIS.2.1.011006},
archivePrefix = {arXiv},
       eprint = {1511.02869},
 primaryClass = {astro-ph.IM},
       adsurl = {https://ui.adsabs.harvard.edu/abs/2016JATIS...2a1006S}
}

@ARTICLE{Spergel2015arXiv150303757S,
       author = {{Spergel}, D. and {Gehrels}, N. and {Baltay}, C. and {Bennett}, D. and {Breckinridge}, J. and {Donahue}, M. and {Dressler}, A. and {Gaudi}, B.~S. and {Greene}, T. and {Guyon}, O. and {Hirata}, C. and {Kalirai}, J. and {Kasdin}, N.~J. and {Macintosh}, B. and {Moos}, W. and {Perlmutter}, S. and {Postman}, M. and {Rauscher}, B. and {Rhodes}, J. and {Wang}, Y. and {Weinberg}, D. and {Benford}, D. and {Hudson}, M. and {Jeong}, W. -S. and {Mellier}, Y. and {Traub}, W. and {Yamada}, T. and {Capak}, P. and {Colbert}, J. and {Masters}, D. and {Penny}, M. and {Savransky}, D. and {Stern}, D. and {Zimmerman}, N. and {Barry}, R. and {Bartusek}, L. and {Carpenter}, K. and {Cheng}, E. and {Content}, D. and {Dekens}, F. and {Demers}, R. and {Grady}, K. and {Jackson}, C. and {Kuan}, G. and {Kruk}, J. and {Melton}, M. and {Nemati}, B. and {Parvin}, B. and {Poberezhskiy}, I. and {Peddie}, C. and {Ruffa}, J. and {Wallace}, J.~K. and {Whipple}, A. and {Wollack}, E. and {Zhao}, F.},
        title = "{Wide-Field InfrarRed Survey Telescope-Astrophysics Focused Telescope Assets WFIRST-AFTA 2015 Report}",
      journal = {arXiv e-prints},
         year = 2015,
        month = mar,
          eid = {arXiv:1503.03757},
        pages = {arXiv:1503.03757},
          doi = {10.48550/arXiv.1503.03757},
archivePrefix = {arXiv},
       eprint = {1503.03757},
 primaryClass = {astro-ph.IM},
       adsurl = {https://ui.adsabs.harvard.edu/abs/2015arXiv150303757S}
}

@ARTICLE{Ruffio2017ApJ...842...14R,
       author = {{Ruffio}, Jean-Baptiste and {Macintosh}, Bruce and {Wang}, Jason J. and {Pueyo}, Laurent and {Nielsen}, Eric L. and {De Rosa}, Robert J. and {Czekala}, Ian and {Marley}, Mark S. and {Arriaga}, Pauline and {Bailey}, Vanessa P. and {Barman}, Travis and {Bulger}, Joanna and {Chilcote}, Jeffrey and {Cotten}, Tara and {Doyon}, Rene and {Duch{\^e}ne}, Gaspard and {Fitzgerald}, Michael P. and {Follette}, Katherine B. and {Gerard}, Benjamin L. and {Goodsell}, Stephen J. and {Graham}, James R. and {Greenbaum}, Alexandra Z. and {Hibon}, Pascale and {Hung}, Li-Wei and {Ingraham}, Patrick and {Kalas}, Paul and {Konopacky}, Quinn and {Larkin}, James E. and {Maire}, J{\'e}r{\^o}me and {Marchis}, Franck and {Marois}, Christian and {Metchev}, Stanimir and {Millar-Blanchaer}, Maxwell A. and {Morzinski}, Katie M. and {Oppenheimer}, Rebecca and {Palmer}, David and {Patience}, Jennifer and {Perrin}, Marshall and {Poyneer}, Lisa and {Rajan}, Abhijith and {Rameau}, Julien and {Rantakyr{\"o}}, Fredrik T. and {Savransky}, Dmitry and {Schneider}, Adam C. and {Sivaramakrishnan}, Anand and {Song}, Inseok and {Soummer}, Remi and {Thomas}, Sandrine and {Wallace}, J. Kent and {Ward-Duong}, Kimberly and {Wiktorowicz}, Sloane and {Wolff}, Schuyler},
        title = "{Improving and Assessing Planet Sensitivity of the GPI Exoplanet Survey with a Forward Model Matched Filter}",
      journal = {\apj},
         year = 2017,
        month = jun,
       volume = {842},
       number = {1},
          eid = {14},
        pages = {14},
          doi = {10.3847/1538-4357/aa72dd},
archivePrefix = {arXiv},
       eprint = {1705.05477},
 primaryClass = {astro-ph.EP},
       adsurl = {https://ui.adsabs.harvard.edu/abs/2017ApJ...842...14R}
}

@ARTICLE{Landman2023A&A...675A.157L,
       author = {{Landman}, R. and {Snellen}, I.~A.~G. and {Keller}, C.~U. and {N'Diaye}, M. and {Fagginger-Auer}, F. and {Desgrange}, C.},
        title = "{Trade-offs in high-contrast integral field spectroscopy for exoplanet detection and characterisation. Young gas giants in emission}",
      journal = {\aap},
         year = 2023,
        month = jul,
       volume = {675},
          eid = {A157},
        pages = {A157},
          doi = {10.1051/0004-6361/202245169},
archivePrefix = {arXiv},
       eprint = {2305.19355},
 primaryClass = {astro-ph.EP},
       adsurl = {https://ui.adsabs.harvard.edu/abs/2023A&A...675A.157L}
}

@ARTICLE{Ruffio2019AJ....158..200R,
       author = {{Ruffio}, Jean-Baptiste and {Macintosh}, Bruce and {Konopacky}, Quinn M. and {Barman}, Travis and {De Rosa}, Robert J. and {Wang}, Jason J. and {Wilcomb}, Kielan K. and {Czekala}, Ian and {Marois}, Christian},
        title = "{Radial Velocity Measurements of HR 8799 b and c with Medium Resolution Spectroscopy}",
      journal = {\aj},
         year = 2019,
        month = nov,
       volume = {158},
       number = {5},
          eid = {200},
        pages = {200},
          doi = {10.3847/1538-3881/ab4594},
archivePrefix = {arXiv},
       eprint = {1909.07571},
 primaryClass = {astro-ph.EP},
       adsurl = {https://ui.adsabs.harvard.edu/abs/2019AJ....158..200R}
}

@ARTICLE{Xuan2024ApJ...970...71X,
       author = {{Xuan}, Jerry W. and {Hsu}, Chih-Chun and {Finnerty}, Luke and {Wang}, Jason and {Ruffio}, Jean-Baptiste and {Zhang}, Yapeng and {Knutson}, Heather A. and {Mawet}, Dimitri and {Mamajek}, Eric E. and {Inglis}, Julie and {Wallack}, Nicole L. and {Bryan}, Marta L. and {Blake}, Geoffrey A. and {Molli{\`e}re}, Paul and {Hejazi}, Neda and {Baker}, Ashley and {Bartos}, Randall and {Calvin}, Benjamin and {Cetre}, Sylvain and {Delorme}, Jacques-Robert and {Doppmann}, Greg and {Echeverri}, Daniel and {Fitzgerald}, Michael P. and {Jovanovic}, Nemanja and {Liberman}, Joshua and {L{\'o}pez}, Ronald A. and {Morris}, Evan and {Pezzato}, Jacklyn and {Sappey}, Ben and {Schofield}, Tobias and {Skemer}, Andrew and {Wallace}, J. Kent and {Wang}, Ji and {Agrawal}, Shubh and {Horstman}, Katelyn},
        title = "{Are These Planets or Brown Dwarfs? Broadly Solar Compositions from High-resolution Atmospheric Retrievals of {\ensuremath{\sim}}10{\textendash}30 M $_{Jup}$ Companions}",
      journal = {\apj},
         year = 2024,
        month = jul,
       volume = {970},
       number = {1},
          eid = {71},
        pages = {71},
          doi = {10.3847/1538-4357/ad4796},
archivePrefix = {arXiv},
       eprint = {2405.13128},
 primaryClass = {astro-ph.EP},
       adsurl = {https://ui.adsabs.harvard.edu/abs/2024ApJ...970...71X}
}

@ARTICLE{Hoch2023AJ....166...85H,
       author = {{Hoch}, Kielan K.~W. and {Konopacky}, Quinn M. and {Theissen}, Christopher A. and {Ruffio}, Jean-Baptiste and {Barman}, Travis S. and {Rickman}, Emily L. and {Perrin}, Marshall D. and {Macintosh}, Bruce and {Marois}, Christian},
        title = "{Assessing the C/O Ratio Formation Diagnostic: A Potential Trend with Companion Mass}",
      journal = {\aj},
         year = 2023,
        month = sep,
       volume = {166},
       number = {3},
          eid = {85},
        pages = {85},
          doi = {10.3847/1538-3881/ace442},
archivePrefix = {arXiv},
       eprint = {2212.04557},
 primaryClass = {astro-ph.EP},
       adsurl = {https://ui.adsabs.harvard.edu/abs/2023AJ....166...85H}
}

@INPROCEEDINGS{Nemati2014SPIE.9143E..0QN,
       author = {{Nemati}, Bijan},
        title = "{Detector selection for the WFIRST-AFTA coronagraph integral field spectrograph}",
    booktitle = {Space Telescopes and Instrumentation 2014: Optical, Infrared, and Millimeter Wave},
         year = 2014,
       editor = {{Oschmann}, Jr., Jacobus M. and {Clampin}, Mark and {Fazio}, Giovanni G. and {MacEwen}, Howard A.},
       series = {Society of Photo-Optical Instrumentation Engineers (SPIE) Conference Series},
       volume = {9143},
        month = aug,
          eid = {91430Q},
        pages = {91430Q},
          doi = {10.1117/12.2060321},
       adsurl = {https://ui.adsabs.harvard.edu/abs/2014SPIE.9143E..0QN}
}

@ARTICLE{Stark2025arXiv250218556S,
       author = {{Stark}, Christopher C. and {Steiger}, Sarah and {Tokadjian}, Armen and {Savransky}, Dmitry and {Belikov}, Rus and {Chen}, Pin and {Krist}, John and {Macintosh}, Bruce and {Morgan}, Rhonda and {Pueyo}, Laurent and {Sirbu}, Dan and {Stapelfeldt}, Karl},
        title = "{Cross-Model Validation of Coronagraphic Exposure Time Calculators for the Habitable Worlds Observatory: A Report from the Exoplanet Science Yield sub-Working Group}",
      journal = {arXiv e-prints},
         year = 2025,
        month = feb,
          eid = {arXiv:2502.18556},
        pages = {arXiv:2502.18556},
          doi = {10.48550/arXiv.2502.18556},
archivePrefix = {arXiv},
       eprint = {2502.18556},
 primaryClass = {astro-ph.IM},
       adsurl = {https://ui.adsabs.harvard.edu/abs/2025arXiv250218556S}
}

@ARTICLE{Ertel2020AJ....159..177E,
       author = {{Ertel}, S. and {Defr{\`e}re}, D. and {Hinz}, P. and {Mennesson}, B. and {Kennedy}, G.~M. and {Danchi}, W.~C. and {Gelino}, C. and {Hill}, J.~M. and {Hoffmann}, W.~F. and {Mazoyer}, J. and {Rieke}, G. and {Shannon}, A. and {Stapelfeldt}, K. and {Spalding}, E. and {Stone}, J.~M. and {Vaz}, A. and {Weinberger}, A.~J. and {Willems}, P. and {Absil}, O. and {Arbo}, P. and {Bailey}, V.~P. and {Beichman}, C. and {Bryden}, G. and {Downey}, E.~C. and {Durney}, O. and {Esposito}, S. and {Gaspar}, A. and {Grenz}, P. and {Haniff}, C.~A. and {Leisenring}, J.~M. and {Marion}, L. and {McMahon}, T.~J. and {Millan-Gabet}, R. and {Montoya}, M. and {Morzinski}, K.~M. and {Perera}, S. and {Pinna}, E. and {Pott}, J. -U. and {Power}, J. and {Puglisi}, A. and {Roberge}, A. and {Serabyn}, E. and {Skemer}, A.~J. and {Su}, K.~Y.~L. and {Vaitheeswaran}, V. and {Wyatt}, M.~C.},
        title = "{The HOSTS Survey for Exozodiacal Dust: Observational Results from the Complete Survey}",
      journal = {\aj},
         year = 2020,
        month = apr,
       volume = {159},
       number = {4},
          eid = {177},
        pages = {177},
          doi = {10.3847/1538-3881/ab7817},
archivePrefix = {arXiv},
       eprint = {2003.03499},
 primaryClass = {astro-ph.SR},
       adsurl = {https://ui.adsabs.harvard.edu/abs/2020AJ....159..177E}
}

@ARTICLE{Wang2017AJ....153..183W,
       author = {{Wang}, Ji and {Mawet}, Dimitri and {Ruane}, Garreth and {Hu}, Renyu and {Benneke}, Bj{\"o}rn},
        title = "{Observing Exoplanets with High Dispersion Coronagraphy. I. The Scientific Potential of Current and Next-generation Large Ground and Space Telescopes}",
      journal = {\aj},
         year = 2017,
        month = apr,
       volume = {153},
       number = {4},
          eid = {183},
        pages = {183},
          doi = {10.3847/1538-3881/aa6474},
archivePrefix = {arXiv},
       eprint = {1703.00582},
 primaryClass = {astro-ph.EP},
       adsurl = {https://ui.adsabs.harvard.edu/abs/2017AJ....153..183W}
}

@ARTICLE{Boker2022,
       author = {{B{\"o}ker}, T. and {Arribas}, S. and {L{\"u}tzgendorf}, N. and {Alves de Oliveira}, C. and {Beck}, T.~L. and {Birkmann}, S. and {Bunker}, A.~J. and {Charlot}, S. and {de Marchi}, G. and {Ferruit}, P. and {Giardino}, G. and {Jakobsen}, P. and {Kumari}, N. and {L{\'o}pez-Caniego}, M. and {Maiolino}, R. and {Manjavacas}, E. and {Marston}, A. and {Moseley}, S.~H. and {Muzerolle}, J. and {Ogle}, P. and {Pirzkal}, N. and {Rauscher}, B. and {Rawle}, T. and {Rix}, H. -W. and {Sabbi}, E. and {Sargent}, B. and {Sirianni}, M. and {te Plate}, M. and {Valenti}, J. and {Willott}, C.~J. and {Zeidler}, P.},
        title = "{The Near-Infrared Spectrograph (NIRSpec) on the James Webb Space Telescope. III. Integral-field spectroscopy}",
      journal = {Astron. Astrophys.},
         year = 2022,
        month = may,
       volume = {661},
          eid = {A82},
        pages = {A82},
          doi = {10.1051/0004-6361/202142589},
archivePrefix = {arXiv},
       eprint = {2202.03308},
 primaryClass = {astro-ph.IM},
       adsurl = {https://ui.adsabs.harvard.edu/abs/2022A&A...661A..82B}
}

@ARTICLE{Gaudi2020arXiv200106683G,
       author = {{Gaudi}, B. Scott and {Seager}, Sara and {Mennesson}, Bertrand and {Kiessling}, Alina and {Warfield}, Keith and {Cahoy}, Kerri and {Clarke}, John T. and {Domagal-Goldman}, Shawn and {Feinberg}, Lee and {Guyon}, Olivier and {Kasdin}, Jeremy and {Mawet}, Dimitri and {Plavchan}, Peter and {Robinson}, Tyler and {Rogers}, Leslie and {Scowen}, Paul and {Somerville}, Rachel and {Stapelfeldt}, Karl and {Stark}, Christopher and {Stern}, Daniel and {Turnbull}, Margaret and {Amini}, Rashied and {Kuan}, Gary and {Martin}, Stefan and {Morgan}, Rhonda and {Redding}, David and {Stahl}, H. Philip and {Webb}, Ryan and {Alvarez-Salazar}, Oscar and {Arnold}, William L. and {Arya}, Manan and {Balasubramanian}, Bala and {Baysinger}, Mike and {Bell}, Ray and {Below}, Chris and {Benson}, Jonathan and {Blais}, Lindsey and {Booth}, Jeff and {Bourgeois}, Robert and {Bradford}, Case and {Brewer}, Alden and {Brooks}, Thomas and {Cady}, Eric and {Caldwell}, Mary and {Calvet}, Rob and {Carr}, Steven and {Chan}, Derek and {Cormarkovic}, Velibor and {Coste}, Keith and {Cox}, Charlie and {Danner}, Rolf and {Davis}, Jacqueline and {Dewell}, Larry and {Dorsett}, Lisa and {Dunn}, Daniel and {East}, Matthew and {Effinger}, Michael and {Eng}, Ron and {Freebury}, Greg and {Garcia}, Jay and {Gaskin}, Jonathan and {Greene}, Suzan and {Hennessy}, John and {Hilgemann}, Evan and {Hood}, Brad and {Holota}, Wolfgang and {Howe}, Scott and {Huang}, Pei and {Hull}, Tony and {Hunt}, Ron and {Hurd}, Kevin and {Johnson}, Sandra and {Kissil}, Andrew and {Knight}, Brent and {Kolenz}, Daniel and {Kraus}, Oliver and {Krist}, John and {Li}, Mary and {Lisman}, Doug and {Mandic}, Milan and {Mann}, John and {Marchen}, Luis and {Marrese-Reading}, Colleen and {McCready}, Jonathan and {McGown}, Jim and {Missun}, Jessica and {Miyaguchi}, Andrew and {Moore}, Bradley and {Nemati}, Bijan and {Nikzad}, Shouleh and {Nissen}, Joel and {Novicki}, Megan and {Perrine}, Todd and {Pineda}, Claudia and {Polanco}, Otto and {Putnam}, Dustin and {Qureshi}, Atif and {Richards}, Michael and {Eldorado Riggs}, A.~J. and {Rodgers}, Michael and {Rud}, Mike and {Saini}, Navtej and {Scalisi}, Dan and {Scharf}, Dan and {Schulz}, Kevin and {Serabyn}, Gene and {Sigrist}, Norbert and {Sikkia}, Glory and {Singleton}, Andrew and {Shaklan}, Stuart and {Smith}, Scott and {Southerd}, Bart and {Stahl}, Mark and {Steeves}, John and {Sturges}, Brian and {Sullivan}, Chris and {Tang}, Hao and {Taras}, Neil and {Tesch}, Jonathan and {Therrell}, Melissa and {Tseng}, Howard and {Valente}, Marty and {Van Buren}, David and {Villalvazo}, Juan and {Warwick}, Steve and {Webb}, David and {Westerhoff}, Thomas and {Wofford}, Rush and {Wu}, Gordon and {Woo}, Jahning and {Wood}, Milana and {Ziemer}, John and {Arney}, Giada and {Anderson}, Jay and {Ma{\'\i}z-Apell{\'a}niz}, Jes{\'u}s and {Bartlett}, James and {Belikov}, Ruslan and {Bendek}, Eduardo and {Cenko}, Brad and {Douglas}, Ewan and {Dulz}, Shannon and {Evans}, Chris and {Faramaz}, Virginie and {Feng}, Y. Katherina and {Ferguson}, Harry and {Follette}, Kate and {Ford}, Saavik and {Garc{\'\i}a}, Miriam and {Geha}, Marla and {Gelino}, Dawn and {G{\"o}tberg}, Ylva and {Hildebrandt}, Sergi and {Hu}, Renyu and {Jahnke}, Knud and {Kennedy}, Grant and {Kreidberg}, Laura and {Isella}, Andrea and {Lopez}, Eric and {Marchis}, Franck and {Macri}, Lucas and {Marley}, Mark and {Matzko}, William and {Mazoyer}, Johan and {McCandliss}, Stephan and {Meshkat}, Tiffany and {Mordasini}, Christoph and {Morris}, Patrick and {Nielsen}, Eric and {Newman}, Patrick and {Petigura}, Erik and {Postman}, Marc and {Reines}, Amy and {Roberge}, Aki and {Roederer}, Ian and {Ruane}, Garreth and {Schwieterman}, Edouard and {Sirbu}, Dan and {Spalding}, Christopher and {Teplitz}, Harry and {Tumlinson}, Jason and {Turner}, Neal and {Werk}, Jessica and {Wofford}, Aida and {Wyatt}, Mark and {Young}, Amber and {Zellem}, Rob},
        title = "{The Habitable Exoplanet Observatory (HabEx) Mission Concept Study Final Report}",
      journal = {arXiv e-prints},
         year = 2020,
        month = jan,
          eid = {arXiv:2001.06683},
        pages = {arXiv:2001.06683},
          doi = {10.48550/arXiv.2001.06683},
archivePrefix = {arXiv},
       eprint = {2001.06683},
 primaryClass = {astro-ph.IM},
       adsurl = {https://ui.adsabs.harvard.edu/abs/2020arXiv200106683G}
}

@ARTICLE{LUVOIR2019arXiv191206219T,
       author = {{The LUVOIR Team}},
        title = "{The LUVOIR Mission Concept Study Final Report}",
      journal = {arXiv e-prints},
         year = 2019,
        month = dec,
          eid = {arXiv:1912.06219},
        pages = {arXiv:1912.06219},
          doi = {10.48550/arXiv.1912.06219},
archivePrefix = {arXiv},
       eprint = {1912.06219},
 primaryClass = {astro-ph.IM},
       adsurl = {https://ui.adsabs.harvard.edu/abs/2019arXiv191206219T}
}

@INPROCEEDINGS{Horstman2024SPIE13096E..2EH,
       author = {{Horstman}, Katelyn A. and {Ruffio}, Jean-Baptiste and {Wang}, Jason J. and {Hsu}, Chih-Chun and {Baker}, Ashley and {Finnerty}, Luke and {Xuan}, Jerry and {Echeverri}, Daniel and {Mawet}, Dimitri and {Blake}, Geoffrey A. and {Bartos}, Randall and {Bond}, Charlotte Z. and {Calvin}, Benjamin and {Cetre}, Sylvain and {Delorme}, Jacques-Robert and {Doppmann}, Greg and {Fitzgerald}, Michael P. and {Jovanovic}, Nemanja J. and {Lopez}, Ronald and {Martin}, Emily C. and {Morris}, Evan and {Pezzato}, Jacklyn and {Ruane}, Garreth and {Sappey}, Ben and {Schofield}, Tobias and {Skemer}, Andrew and {Venenciano}, Taylor and {Wallace}, J. Kent and {Wang}, Ji and {Wizinowich}, Peter},
        title = "{Fringing analysis and forward modeling of Keck Planet Imager and Characterizer (KPIC) spectra}",
    booktitle = {Ground-based and Airborne Instrumentation for Astronomy X},
         year = 2024,
       editor = {{Bryant}, Julia J. and {Motohara}, Kentaro and {Vernet}, Jo{\"e}l. R.~D.},
       series = {Society of Photo-Optical Instrumentation Engineers (SPIE) Conference Series},
       volume = {13096},
        month = jul,
          eid = {130962E},
        pages = {130962E},
          doi = {10.1117/12.3018020},
       adsurl = {https://ui.adsabs.harvard.edu/abs/2024SPIE13096E..2EH}
}

@ARTICLE{Gasman2023A&A...673A.102G,
       author = {{Gasman}, Danny and {Argyriou}, Ioannis and {Sloan}, G.~C. and {Aringer}, Bernhard and {{\'A}lvarez-M{\'a}rquez}, Javier and {Fox}, Ori and {Glasse}, Alistair and {Glauser}, Adrian and {Jones}, Olivia C. and {Justtanont}, Kay and {Kavanagh}, Patrick J. and {Klaassen}, Pamela and {Labiano}, Alvaro and {Larson}, Kirsten and {Law}, David R. and {Mueller}, Michael and {Nayak}, Omnarayani and {Noriega-Crespo}, Alberto and {Patapis}, Polychronis and {Royer}, Pierre and {Vandenbussche}, Bart},
        title = "{JWST MIRI/MRS in-flight absolute flux calibration and tailored fringe correction for unresolved sources}",
      journal = {\aap},
         year = 2023,
        month = may,
       volume = {673},
          eid = {A102},
        pages = {A102},
          doi = {10.1051/0004-6361/202245633},
archivePrefix = {arXiv},
       eprint = {2212.03596},
 primaryClass = {astro-ph.IM},
       adsurl = {https://ui.adsabs.harvard.edu/abs/2023A&A...673A.102G}
}

@ARTICLE{McElwain2023PASP..135e8001M,
       author = {{McElwain}, Michael W. and {Feinberg}, Lee D. and {Perrin}, Marshall D. and {Clampin}, Mark and {Mountain}, C. Matt and {Lallo}, Matthew D. and {Lajoie}, Charles-Philippe and {Kimble}, Randy A. and {Bowers}, Charles W. and {Stark}, Christopher C. and {Acton}, D. Scott and {Atkinson}, Charles and {Barinek}, Beth and {Barto}, Allison and {Basinger}, Scott and {Beck}, Tracy and {Bergkoetter}, Matthew D. and {Bluth}, Marcel and {Boucarut}, Rene A. and {Brady}, Gregory R. and {Brooks}, Keira J. and {Brown}, Bob and {Byard}, John and {Carey}, Larkin and {Carrasquilla}, Maria and {Chae}, Dan and {Chaney}, David and {Chayer}, Pierre and {Chonis}, Taylor and {Cohen}, Lester and {Cole}, Helen J. and {Comeau}, Thomas M. and {Coon}, Matthew and {Coppock}, Eric and {Coyle}, Laura and {Dean}, Bruce H. and {Dziak}, Kenneth J. and {Eisenhower}, Michael and {Flagey}, Nicolas and {Franck}, Randy and {Gallagher}, Benjamin and {Gilman}, Larry and {Glassman}, Tiffany and {Green}, Joseph J. and {Grieco}, John and {Haase}, Shari and {Hadjimichael}, Theodore J. and {Hagopian}, John G. and {Hahn}, Walter G. and {Hartig}, George F. and {Havey}, Keith A. and {Hayden}, William L. and {Hellekson}, Robert and {Hicks}, Brian and {Holfeltz}, Sherie T. and {Howard}, Joseph M. and {Huguet}, Jesse A. and {Jahne}, Brian and {Johnson}, Leslie A. and {Johnston}, John D. and {Jurling}, Alden S. and {Kegley}, Jeffrey R. and {Kennard}, Scott and {Keski-Kuha}, Ritva A. and {Knight}, J. Scott and {Kulp}, Bernard A. and {Levi}, Joshua S. and {Levine}, Marie B. and {Lightsey}, Paul and {Luetgens}, Robert A. and {Mather}, John C. and {Matthews}, Gary W. and {McKay}, Andrew G. and {Mehalick}, Kimberly I. and {Mel{\'e}ndez}, Marcio and {Mosier}, Gary E. and {Murphy}, Jess and {Nelan}, Edmund P. and {Niedner}, Malcolm B. and {Nol}, Darin M. and {Ohara}, Catherine M. and {Ohl}, Raymond G. and {Olczak}, Eugene and {Osborne}, Shannon B. and {Park}, Sang and {Perrygo}, Charles and {Pueyo}, Laurent and {Redding}, David C. and {Regan}, Michael W. and {Reynolds}, Paul and {Rifelli}, Rich and {Rigby}, Jane R. and {Sabatke}, Derek and {Saif}, Babak N. and {Scorse}, Thomas R. and {Seo}, Byoung-Joon and {Shi}, Fang and {Sigrist}, Norbert and {Smith}, Koby and {Smith}, J. Scott and {Smith}, Erin C. and {Sohn}, Sangmo Tony and {Stahl}, H. Philip and {Telfer}, Randal and {Terlecki}, Todd and {Texter}, Scott C. and {Van Buren}, David and {Van Campen}, Julie M. and {Vila}, Bego{\~n}a and {Voyton}, Mark F. and {Waldman}, Mark and {Walker}, Chanda B. and {Weiser}, Nick and {Wells}, Conrad and {West}, Garrett and {Whitman}, Tony L. and {Wolf}, Erin and {Zielinski}, Thomas P.},
        title = "{The James Webb Space Telescope Mission: Optical Telescope Element Design, Development, and Performance}",
      journal = {\pasp},
         year = 2023,
        month = may,
       volume = {135},
       number = {1047},
          eid = {058001},
        pages = {058001},
          doi = {10.1088/1538-3873/acada0},
archivePrefix = {arXiv},
       eprint = {2301.01779},
 primaryClass = {astro-ph.IM},
       adsurl = {https://ui.adsabs.harvard.edu/abs/2023PASP..135e8001M}
}

@ARTICLE{Konopacky2013Sci...339.1398K,
       author = {{Konopacky}, Quinn M. and {Barman}, Travis S. and {Macintosh}, Bruce A. and {Marois}, Christian},
        title = "{Detection of Carbon Monoxide and Water Absorption Lines in an Exoplanet Atmosphere}",
      journal = {Science},
         year = 2013,
        month = mar,
       volume = {339},
       number = {6126},
        pages = {1398-1401},
          doi = {10.1126/science.1232003},
archivePrefix = {arXiv},
       eprint = {1303.3280},
 primaryClass = {astro-ph.EP},
       adsurl = {https://ui.adsabs.harvard.edu/abs/2013Sci...339.1398K}
}

@ARTICLE{Snellen2025arXiv250508926S,
       author = {{Snellen}, Ignas},
        title = "{Exoplanet atmospheres at high spectral resolution}",
      journal = {arXiv e-prints},
         year = 2025,
        month = may,
          eid = {arXiv:2505.08926},
        pages = {arXiv:2505.08926},
          doi = {10.48550/arXiv.2505.08926},
archivePrefix = {arXiv},
       eprint = {2505.08926},
 primaryClass = {astro-ph.EP},
       adsurl = {https://ui.adsabs.harvard.edu/abs/2025arXiv250508926S}
}

@ARTICLE{Wang2018JATIS...4c5001W,
       author = {{Wang}, Ji and {Mawet}, Dimitri and {Hu}, Renyu and {Ruane}, Garreth and {Delorme}, Jacques-Robert and {Klimovich}, Nikita},
        title = "{Baseline requirements for detecting biosignatures with the HabEx and LUVOIR mission concepts}",
      journal = {Journal of Astronomical Telescopes, Instruments, and Systems},
         year = 2018,
        month = jul,
       volume = {4},
          eid = {035001},
        pages = {035001},
          doi = {10.1117/1.JATIS.4.3.035001},
archivePrefix = {arXiv},
       eprint = {1806.04324},
 primaryClass = {astro-ph.EP},
       adsurl = {https://ui.adsabs.harvard.edu/abs/2018JATIS...4c5001W}
}

@ARTICLE{Hoeijmakers2018A&A...617A.144H,
       author = {{Hoeijmakers}, H.~J. and {Schwarz}, H. and {Snellen}, I.~A.~G. and {de Kok}, R.~J. and {Bonnefoy}, M. and {Chauvin}, G. and {Lagrange}, A.~M. and {Girard}, J.~H.},
        title = "{Medium-resolution integral-field spectroscopy for high-contrast exoplanet imaging. Molecule maps of the {\ensuremath{\beta}} Pictoris system with SINFONI}",
      journal = {\aap},
         year = 2018,
        month = oct,
       volume = {617},
          eid = {A144},
        pages = {A144},
          doi = {10.1051/0004-6361/201832902},
archivePrefix = {arXiv},
       eprint = {1802.09721},
 primaryClass = {astro-ph.EP},
       adsurl = {https://ui.adsabs.harvard.edu/abs/2018A&A...617A.144H}
}

@ARTICLE{Zhang2024PASP..136e4401Z,
       author = {{Zhang}, Jingwen and {Bottom}, Michael and {Serabyn}, Eugene},
        title = "{Direct Detection and Characterization of Exoplanets Using Imaging Fourier Transform Spectroscopy}",
      journal = {\pasp},
         year = 2024,
        month = may,
       volume = {136},
       number = {5},
          eid = {054401},
        pages = {054401},
          doi = {10.1088/1538-3873/ad37d8},
archivePrefix = {arXiv},
       eprint = {2310.15231},
 primaryClass = {astro-ph.EP},
       adsurl = {https://ui.adsabs.harvard.edu/abs/2024PASP..136e4401Z}
}

@INPROCEEDINGS{Wang2017SPIE10400E..0ZW,
       author = {{Wang}, Ji and {Mawet}, Dimitri and {Ruane}, Garreth and {Delorme}, Jacques-Robert and {Klimovich}, Nikita and {Hu}, Renyu},
        title = "{Baseline requirements for detecting biosignatures with the HabEx and LUVOIR mission concepts}",
    booktitle = {Society of Photo-Optical Instrumentation Engineers (SPIE) Conference Series},
         year = 2017,
       editor = {{Shaklan}, Stuart},
       series = {Society of Photo-Optical Instrumentation Engineers (SPIE) Conference Series},
       volume = {10400},
        month = sep,
          eid = {104000Z},
        pages = {104000Z},
          doi = {10.1117/12.2275222},
       adsurl = {https://ui.adsabs.harvard.edu/abs/2017SPIE10400E..0ZW}
}

@ARTICLE{Bidot2024A&A...682A..10B,
       author = {{Bidot}, A. and {Mouillet}, D. and {Carlotti}, A.},
        title = "{Exoplanet detection limits using spectral cross-correlation with spectro-imaging. Analytical model applied to the case of ELT/HARMONI}",
      journal = {\aap},
         year = 2024,
        month = feb,
       volume = {682},
          eid = {A10},
        pages = {A10},
          doi = {10.1051/0004-6361/202346185},
archivePrefix = {arXiv},
       eprint = {2311.13275},
 primaryClass = {astro-ph.IM},
       adsurl = {https://ui.adsabs.harvard.edu/abs/2024A&A...682A..10B}
}

@ARTICLE{Hardegree2025AJ....169..171H,
       author = {{Hardegree-Ullman}, Kevin K. and {Apai}, D{\'a}niel and {Haffert}, Sebastiaan Y. and {Schlecker}, Martin and {Kasper}, Markus and {Kammerer}, Jens and {Wagner}, Kevin},
        title = "{Bioverse: Giant Magellan Telescope and Extremely Large Telescope Direct Imaging and High-resolution Spectroscopy Assessment{\textemdash}Surveying Exo-Earth O$_{2}$ and Testing the Habitable Zone Oxygen Hypothesis}",
      journal = {\aj},
         year = 2025,
        month = mar,
       volume = {169},
       number = {3},
          eid = {171},
        pages = {171},
          doi = {10.3847/1538-3881/adb02f},
archivePrefix = {arXiv},
       eprint = {2405.11423},
 primaryClass = {astro-ph.EP},
       adsurl = {https://ui.adsabs.harvard.edu/abs/2025AJ....169..171H}
}

@ARTICLE{Law2023AJ....166...45L,
       author = {{Law}, David R. and {E. Morrison}, Jane and {Argyriou}, Ioannis and {Patapis}, Polychronis and {{\'A}lvarez-M{\'a}rquez}, J. and {Labiano}, Alvaro and {Vandenbussche}, Bart},
        title = "{A 3D Drizzle Algorithm for JWST and Practical Application to the MIRI Medium Resolution Spectrometer}",
      journal = {\aj},
         year = 2023,
        month = aug,
       volume = {166},
       number = {2},
          eid = {45},
        pages = {45},
          doi = {10.3847/1538-3881/acdddc},
archivePrefix = {arXiv},
       eprint = {2306.05520},
 primaryClass = {astro-ph.IM},
       adsurl = {https://ui.adsabs.harvard.edu/abs/2023AJ....166...45L}
}

@INPROCEEDINGS{2024CDS,
       author = {{Belikov}, Ruslan and {Stark}, Christopher and {Siegler}, Nick and {Por}, Emiel and {Mennesson}, Bertrand and {Redmond}, Susan and {Chen}, Pin and {Fogarty}, Kevin and {Guyon}, Olivier and {Juanola-Parramon}, Roser and {Kasdin}, Jeremy and {Krist}, John and {Mawet}, Dimitri and {Morgan}, Rhonda and {Mejia Prada}, Camilo and {Pueyo}, Laurent and {Ruane}, Garreth and {Sirbu}, Dan and {Stapelfeldt}, Karl and {Trauger}, John and {Zimmerman}, Neil and {Alagao}, Mary Angelie M. and {Carlotti}, Alex and {Chafi}, Jamal and {Doleman}, David and {Gersh-Range}, Jessica and {K{\"o}nig}, Lorenzo and {Leboulleux}, Lucille and {Moody}, Dwight and {Riggs}, A.~J. and {Serabyn}, Eugene and {Snik}, Frans and {Wallace}, Kent},
        title = "{Coronagraph design survey for future exoplanet direct imaging space missions}",
    booktitle = {Space Telescopes and Instrumentation 2024: Optical, Infrared, and Millimeter Wave},
         year = 2024,
       editor = {{Coyle}, Laura E. and {Matsuura}, Shuji and {Perrin}, Marshall D.},
       series = {Society of Photo-Optical Instrumentation Engineers (SPIE) Conference Series},
       volume = {13092},
        month = aug,
          eid = {1309266},
        pages = {1309266},
          doi = {10.1117/12.3020614},
       adsurl = {https://ui.adsabs.harvard.edu/abs/2024SPIE13092E..66B}
}

@INPROCEEDINGS{2016SPIE.9911E..19D,
       author = {{Delacroix}, Christian and {Savransky}, Dmitry and {Garrett}, Daniel and {Lowrance}, Patrick and {Morgan}, Rhonda},
        title = "{Science yield modeling with the Exoplanet Open-Source Imaging Mission Simulator (EXOSIMS)}",
    booktitle = {Modeling, Systems Engineering, and Project Management for Astronomy VI},
         year = 2016,
       editor = {{Angeli}, George Z. and {Dierickx}, Philippe},
       series = {Society of Photo-Optical Instrumentation Engineers (SPIE) Conference Series},
       volume = {9911},
        month = aug,
          eid = {991119},
        pages = {991119},
          doi = {10.1117/12.2233913},
       adsurl = {https://ui.adsabs.harvard.edu/abs/2016SPIE.9911E..19D}
}

@INPROCEEDINGS{Feinberg2024SPIE13092E..1NF,
       author = {{Feinberg}, Lee and {Ziemer}, John and {Ansdell}, Megan and {Crooke}, Julie and {Dressing}, Courtney and {Mennesson}, Bertrand and {O'Meara}, John and {Pepper}, Joshua and {Roberge}, Aki},
        title = "{The Habitable Worlds Observatory engineering view: status, plans, and opportunities}",
    booktitle = {Space Telescopes and Instrumentation 2024: Optical, Infrared, and Millimeter Wave},
         year = 2024,
       editor = {{Coyle}, Laura E. and {Matsuura}, Shuji and {Perrin}, Marshall D.},
       series = {Society of Photo-Optical Instrumentation Engineers (SPIE) Conference Series},
       volume = {13092},
        month = aug,
          eid = {130921N},
        pages = {130921N},
          doi = {10.1117/12.3018328},
       adsurl = {https://ui.adsabs.harvard.edu/abs/2024SPIE13092E..1NF}
}

@ARTICLE{Feinberg2026arXiv260111803F,
       author = {{Feinberg}, Lee D. and {Sitarski}, Breann N. and {McElwain}, Michael W. and {Arney}, Giada and {Baker}, Caleb and {Bolcar}, Matthew R. and {Levine}, Marie and {Liu}, Alice and {Mennesson}, Bertrand and {Roberge}, Aki and {Smith}, J. Scott and {Zhao}, Feng and {Ziemer}, John},
        title = "{Habitable Worlds Observatory's Concept and Technology Maturation: Initial Feasibility and Trade Space Exploration}",
      journal = {arXiv e-prints},
         year = 2026,
        month = jan,
          eid = {arXiv:2601.11803},
        pages = {arXiv:2601.11803},
          doi = {10.48550/arXiv.2601.11803},
archivePrefix = {arXiv},
       eprint = {2601.11803},
 primaryClass = {astro-ph.IM},
       adsurl = {https://ui.adsabs.harvard.edu/abs/2026arXiv260111803F}
}

@ARTICLE{Damiano2023AJ....166..157D,
       author = {{Damiano}, Mario and {Hu}, Renyu and {Mennesson}, Bertrand},
        title = "{Reflected Spectroscopy of Small Exoplanets. III. Probing the UV Band to Measure Biosignature Gases}",
      journal = {\aj},
         year = 2023,
        month = oct,
       volume = {166},
       number = {4},
          eid = {157},
        pages = {157},
          doi = {10.3847/1538-3881/acefd3},
archivePrefix = {arXiv},
       eprint = {2308.08490},
 primaryClass = {astro-ph.EP},
       adsurl = {https://ui.adsabs.harvard.edu/abs/2023AJ....166..157D}
}

@ARTICLE{Damiano2022AJ....163..299D,
       author = {{Damiano}, Mario and {Hu}, Renyu},
        title = "{Reflected Spectroscopy of Small Exoplanets II: Characterization of Terrestrial Exoplanets}",
      journal = {\aj},
         year = 2022,
        month = jun,
       volume = {163},
       number = {6},
          eid = {299},
        pages = {299},
          doi = {10.3847/1538-3881/ac6b97},
archivePrefix = {arXiv},
       eprint = {2204.13816},
 primaryClass = {astro-ph.EP},
       adsurl = {https://ui.adsabs.harvard.edu/abs/2022AJ....163..299D}
}

@ARTICLE{Wang2025ApJ...981..138W,
       author = {{Wang}, Ji},
        title = "{Early Accretion of Large Amounts of Solids for Directly Imaged Exoplanets}",
      journal = {\apj},
         year = 2025,
        month = mar,
       volume = {981},
       number = {2},
          eid = {138},
        pages = {138},
          doi = {10.3847/1538-4357/adb42c},
archivePrefix = {arXiv},
       eprint = {2310.00088},
 primaryClass = {astro-ph.EP},
       adsurl = {https://ui.adsabs.harvard.edu/abs/2025ApJ...981..138W}
}

@ARTICLE{Czekala2015,
       author = {{Czekala}, Ian and {Andrews}, Sean M. and {Mandel}, Kaisey S. and {Hogg}, David W. and {Green}, Gregory M.},
        title = "{Constructing a Flexible Likelihood Function for Spectroscopic Inference}",
      journal = {\apj},
         year = 2015,
        month = oct,
       volume = {812},
       number = {2},
          eid = {128},
        pages = {128},
          doi = {10.1088/0004-637X/812/2/128},
archivePrefix = {arXiv},
       eprint = {1412.5177},
 primaryClass = {astro-ph.SR},
       adsurl = {https://ui.adsabs.harvard.edu/abs/2015ApJ...812..128C}
}

@ARTICLE{Flasseur2018A&A...618A.138F,
       author = {{Flasseur}, Olivier and {Denis}, Lo{\"\i}c and {Thi{\'e}baut}, {\'E}ric and {Langlois}, Maud},
        title = "{Exoplanet detection in angular differential imaging by statistical learning of the nonstationary patch covariances. The PACO algorithm}",
      journal = {\aap},
         year = 2018,
        month = oct,
       volume = {618},
          eid = {A138},
        pages = {A138},
          doi = {10.1051/0004-6361/201832745},
       adsurl = {https://ui.adsabs.harvard.edu/abs/2018A&A...618A.138F}
}

@ARTICLE{Howe2024JATIS..10b5008H,
       author = {{Howe}, Alex R. and {Stark}, Christopher C. and {Sadleir}, John E.},
        title = "{Scientific impact of a noiseless energy-resolving detector for a future exoplanet-imaging mission}",
      journal = {Journal of Astronomical Telescopes, Instruments, and Systems},
         year = 2024,
        month = apr,
       volume = {10},
          eid = {025008},
        pages = {025008},
          doi = {10.1117/1.JATIS.10.2.025008},
archivePrefix = {arXiv},
       eprint = {2405.08883},
 primaryClass = {astro-ph.IM},
       adsurl = {https://ui.adsabs.harvard.edu/abs/2024JATIS..10b5008H}
}

@ARTICLE{Agrawal2023AJ....166...15A,
       author = {{Agrawal}, Shubh and {Ruffio}, Jean-Baptiste and {Konopacky}, Quinn M. and {Macintosh}, Bruce and {Mawet}, Dimitri and {Nielsen}, Eric L. and {Hoch}, Kielan K.~W. and {Liu}, Michael C. and {Barman}, Travis S. and {Thompson}, William and {Greenbaum}, Alexandra Z. and {Marois}, Christian and {Patience}, Jenny},
        title = "{Detecting Exoplanets Closer to Stars with Moderate Spectral Resolution Integral-field Spectroscopy}",
      journal = {\aj},
         year = 2023,
        month = jul,
       volume = {166},
       number = {1},
          eid = {15},
        pages = {15},
          doi = {10.3847/1538-3881/acd6a3},
archivePrefix = {arXiv},
       eprint = {2305.10362},
 primaryClass = {astro-ph.EP},
       adsurl = {https://ui.adsabs.harvard.edu/abs/2023AJ....166...15A}
}

@article{SteigerChen2026,
author = {Sarah Steiger and Pin Chen and Laurent Pueyo},
title = {{Incorporating wavefront error, wavefront sensing and control, and sensitivities into exposure time calculations for future space missions with the error budget software}},
volume = {12},
journal = {Journal of Astronomical Telescopes, Instruments, and Systems},
number = {4},
publisher = {SPIE},
pages = {041005},
year = {2026},
doi = {10.1117/1.JATIS.12.4.041005},
URL = {https://doi.org/10.1117/1.JATIS.12.4.041005}
}
\bibliographystyle{spiejour}   

\listoffigures

\subsection*{Biographies}
Jean-Baptiste Ruffio is an assistant research scientist studying extra-solar planets in the Astronomy and Astrophysics department at the University of California, San Diego (UCSD). He has been developing statistical and instrumentation techniques to push the frontiers of planet detection and characterization with the largest telescopes in the world, both on the ground and in space. He obtained a master’s degree from the French Aerospace engineering school (ISAE-Supaero) and a Ph.D. in Physics from Stanford University.

\end{spacing}
\end{document}